\documentclass[english,aip,jcp,amsmath,amssymb,reprint]{revtex4-2}
\usepackage[T1]{fontenc}
\usepackage[utf8]{inputenc}   
\usepackage{color}
\usepackage{babel}
\usepackage{textcomp}
\usepackage{amssymb}
\usepackage{graphicx}
\usepackage{dcolumn}
\usepackage{mciteplus}
\usepackage{amsmath}
\usepackage{physics}
\usepackage{hyperref}
\preprint{AIP/123-QED}
\begin{document}

\title{Full Quantum and Mixed Quantum--Classical Dynamics of Hot Exciton Cooling in Semiconductor Nanocrystals}

\author{Bokang Hou}
\email{bkhou@berkeley.edu}
\affiliation{Department of Chemistry, University of California, Berkeley, California, 94720, United States}
\author{Johan E. Runeson}
\email{johan.runeson@physik.uni-freiburg.de}
\affiliation {Institute of Physics, University of Freiburg, Hermann-Herder-Straße 3, 79104 Freiburg, Germany;}
\author{Samuel L. Rudge}
\affiliation {Institute of Physics, University of Freiburg, Hermann-Herder-Straße 3, 79104 Freiburg, Germany;}
\author{Salvatore Gatto}
\affiliation {Institute of Physics, University of Freiburg, Hermann-Herder-Straße 3, 79104 Freiburg, Germany;}
\author{Hans-Dieter Meyer}
\affiliation {Theoretische Chemie, Physikalisch-Chemisches Institut, Universität Heidelberg, INF 229, D-69120 Heidelberg, Germany;}
\author{Michael Thoss}
\email{michael.thoss@physik.uni-freiburg.de}
\affiliation {Institute of Physics, University of Freiburg, Hermann-Herder-Straße 3, 79104 Freiburg, Germany;}
\author{Eran Rabani}
\email{eran.rabani@mail.huji.ac.il}
\affiliation{Fritz Haber Center for Molecular Dynamics, The Institute of Chemistry and The Institute of Applied Physics, The Hebrew University of Jerusalem, Jerusalem 91904, Israel}
\affiliation{Department of Chemistry, University of California, Berkeley, California, 94720, United States}
\affiliation{Materials Sciences Division, Lawrence Berkeley National Laboratory, Berkeley, California, 94720, United States}

\date{\today}
\begin{abstract}
Hot-exciton relaxation in semiconductor nanocrystals (NCs) is often described using perturbative theories, but their accuracy is difficult to assess for realistic exciton--phonon Hamiltonians. Here, we benchmark the perturbative quantum master equation (QME) and several mixed quantum--classical (MQC) methods against fully quantum mechanical dynamics. Using atomistically parameterized models for CdSe core and CdSe/CdS core--shell NCs, we find that bare CdSe exhibits an ultrafast initial decay followed by slower cooling, whereas the core--shell system is dominated by the slower component. Analysis of reduced models shows that the ultrafast component arises from rapid diabatic state mixing driven by thermal fluctuations of low-frequency phonons, rather than from nuclear-assisted energy relaxation. The QME captures the initial fast decay but can fail for the slower relaxation in the diabatic representation, while the mapping approach to surface hopping (MASH) gives the most consistent agreement with both benchmark dynamics and equilibrium populations. These results establish a benchmark for exciton-cooling dynamics in NCs and clarify the physical regimes in which widely used approximate methods are reliable.
\end{abstract}

\maketitle



\section{Introduction}
Semiconductor nanocrystals (NCs) offer a versatile platform for exploring excited-state carrier dynamics at the nanoscale, which play a central role in determining the optoelectronic behavior and device performance~\cite{talapin2010prospects,garcia2021semiconductor,lin2024colloidal}. Upon absorbing a photon with energy exceeding the optical gap, an NC promotes an electron from the valence band to the conduction band, thereby creating an interacting electron--hole pair, or exciton. The initially prepared exciton state can lie hundreds of meV above the optical band edge and is commonly referred to as a hot exciton. The hot exciton subsequently relaxes toward the band edge, dissipating its excess energy as heat via coupling to the lattice vibrations~\cite{melnychuk2021multicarrier}. In photocatalytic and photovoltaic applications, considerable effort has been devoted to identifying NCs that can sustain hot excitons for longer times, so that their excess energy can be harvested before it is dissipated to heat, in order to drive chemical reactions or generate charge flow~\cite{lin2023carrier, cao2025emerging}. 

Ultrafast spectroscopy has shown that hot-exciton relaxation typically occurs on a sub-picosecond timescale, generally too rapidly for efficient energy harvesting~\cite{hendry2006direct, kambhampati2011hot, ghosh2025atomistically}. An early strategy for slowing this relaxation was based on the phonon-bottleneck picture: quantum confinement increases the energy spacing between excitonic states and could, in principle, hinder phonon-assisted cooling when the level spacing exceeds the characteristic phonon energies~\cite{nozik2001spectroscopy}. In practice, however, experiments have shown that hot-exciton cooling in smaller NCs is often not slower and can even be faster, indicating that the simple phonon-bottleneck picture is incomplete~\cite{schaller2005breaking,cooney2007breaking, kilina2009breaking}.  Additional insight comes from charged NCs, in which the hole is removed through external electron injection or doping; in these systems, relaxation of a hot electron is found to be substantially slower than that of a neutral exciton~\cite{pandey2008slow,wang2021spin,sherman2025revealing}. This contrast highlights that exciton cooling is governed not only by coupling to phonons, but also by the correlated electron--hole nature of the excitonic state~\cite{efros1995breaking,jasrasaria2022simulations}.

The breakdown of the simple phonon-bottleneck picture has motivated theoretical work that moves beyond an independent electron--hole framework. Early work often interpreted the lack of a bottleneck in terms of an Auger-like mechanism, in which the electron overcomes a large energy gap by transferring its excess energy to the hole, thereby promoting it to a deeper-lying state~\cite{efros1995breaking,kharchenko1996auger}. While this picture captures an important aspect of the relaxation process, it remains rooted in a single-particle electron--hole basis. More recent theoretical formulations recast the problem directly in the exciton basis, where hot-exciton cooling is described as relaxation between correlated excitonic states~\cite{jasrasaria2021interplay,jasrasaria2022simulations}. In this representation, the fast relaxation can be mediated by multiphonon processes, in which the energy mismatch is bridged by the emission of multiple vibrational quanta~\cite{jasrasaria2023circumventing}. This excitonic description naturally combines electron--hole correlation with multiphonon relaxation pathways and provides a consistent explanation for the ultrafast hot-exciton cooling timescales observed experimentally~\cite{brosseau2023ultrafast,ghosh2025atomistically}.

Describing exciton cooling with real-time quantum dynamics requires dynamical methods whose regimes of validity are understood. Since fully \textit{ab initio} exciton--phonon Hamiltonians are too costly for routine fully quantum propagation, interpretations of the relaxation dynamics often rely on approximate methods that have not been systematically benchmarked for this problem. Here, we revisit this problem from a real-time quantum dynamics perspective by treating both the excitonic and nuclear degrees of freedom (DOFs) explicitly within an atomistically parameterized model Hamiltonian. Using the thermofield multi-layer multiconfiguration time-dependent Hartree (ML-MCTDH) method as a benchmark, together with an accurate long-time estimate of the equilibrium excitonic populations based on a path integral Monte Carlo (PIMC) approach, we evaluate the performance of several widely used approximate methods, including a perturbative quantum master equation (QME) with memory and different classes of mixed quantum–classical (MQC) methods. 

We consider two types of NCs: a bare CdSe core NC, which exhibits a broad distribution of exciton--phonon couplings spanning the acoustic to optical range, and a CdSe/CdS core--shell NC, whose coupling is dominated primarily by optical modes. For the bare-core NC, we find that the hot-exciton dynamics exhibits two distinct regimes, an ultrafast component with a timescale on the order of $10$~fs followed by a slower relaxation on the order of $100$~fs. The core--shell NC predominantly displays only the slower timescale.  We show that the ultrafast component is associated with rapid excitonic state mixing induced by low-frequency phonons, while the slower component corresponds to phonon-assisted exciton cooling. To connect these findings with experiments, we also analyze the time-dependent average exciton energy, which provides a basis-independent measure of exciton cooling and reveals the presence of coherent lattice modulation. Overall, our simulations provide a detailed mechanistic picture of hot-exciton relaxation and enable a mechanistic discussion of state mixing, dephasing, and nonadiabatic relaxation in semiconductor NCs.

\section{Model and Methods}
\subsection{Model Hamiltonian for Exciton Cooling}
\begin{figure*}[ht!]
    \centering
    \includegraphics[width=0.8\linewidth]{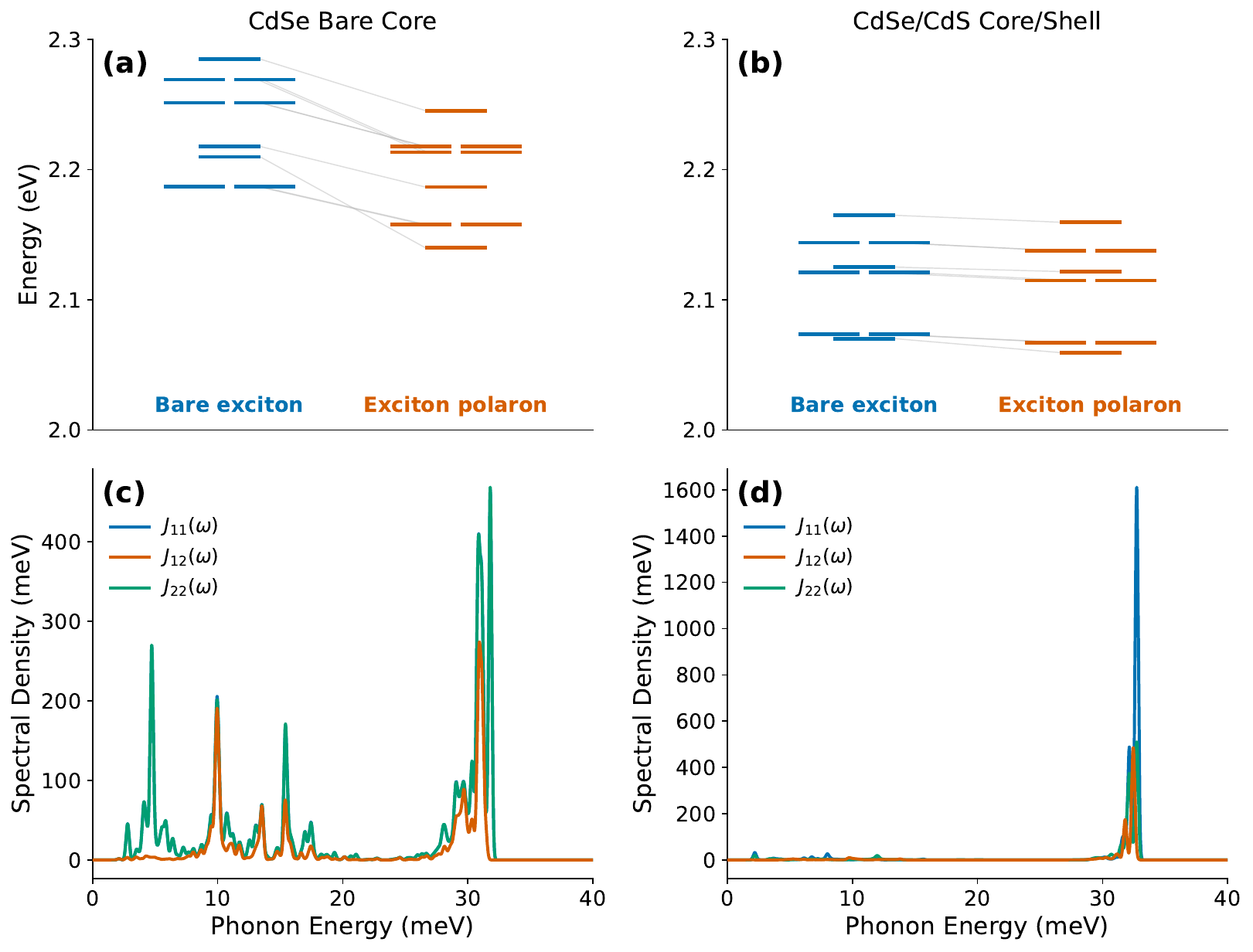}
    \caption{Atomistically parameterized exciton--phonon model for the two NCs.
    (a,b) Bare exciton energies $E_n^0$ at the equilibrium geometry and the corresponding polaron energies $E_n^0-\lambda_{nn}$ for the lowest $9$ excitonic states of (a) the $3$~nm CdSe core NC and (b) the $3$~nm CdSe/ 2ML CdS core--shell NC. (c,d) Exciton--phonon spectral densities $J_{nm}(\omega)$ for the two lowest excitonic states, including the diagonal components \(J_{11}(\omega)\) and \(J_{22}(\omega)\) and the off-diagonal coupling \(J_{12}(\omega)\), for (c) CdSe core and (d) CdSe/CdS core--shell NCs. Note that the $J_{11}(\omega)$ and $J_{22}(\omega)$ for bare-core CdSe overlap. The bare CdSe core shows substantial low-frequency spectral weight below $\sim 25$~meV, whereas the core--shell system is dominated by a narrower optical-phonon feature near $30$--$35$~meV. The spectral densities for higher excitonic states have similar structures.}
    \label{fig:Fig1_level_spec}
\end{figure*}

Simulating NCs of experimentally relevant sizes is challenging because of the large number of excited states involved and the many nuclear degrees of freedom. To reduce the computational complexity, we therefore employ the crude adiabatic representation~\cite{kouppel1984multimode}, model the lattice vibrations within the harmonic approximation, and expand the exciton--nuclear coupling to lowest order in the displacement from equilibrium~\cite{jasrasaria2023circumventing}. The total Hamiltonian can be written as a sum of three terms, $\hat{H}= \hat{H}_\mathrm{ex} + \hat{H}_\mathrm{nu} + \hat{H}_\mathrm{ex-nu}$, where

\begin{equation}
\hat{H}_\mathrm{ex} = \sum_{n=1}^{N_\mathrm{ex}} E_n^0\ket{\psi_n}\bra{\psi_n},
\end{equation}
\begin{equation}
\hat{H}_\mathrm{nu} = \sum_{\alpha=1}^{N_\mathrm{nu}}\left(\frac{\hat{P}_\alpha^2}{2} + \frac{1}{2}\omega_\alpha^2\hat{Q}_\alpha^2\right),
\end{equation}
and
\begin{equation}
\hat{H}_\mathrm{ex-nu} = \sum_{n,m=1}^{N_\mathrm{ex}}\ket{\psi_n}\bra{\psi_m}\sum_{\alpha=1}^{N_\mathrm{nu}}V_{nm}^\alpha \hat{Q}_\alpha.
\end{equation}
The excitonic Hamiltonian $\hat{H}_\mathrm{ex}$ contains the exciton energies $E_n^0$, which are obtained from the Bethe-Salpeter equation based on a semiempirical pseudopotential method~\cite{jasrasaria2022simulations,rabani1999electronic,rohlfing2000electron}. The superscript $0$ indicates that these exciton energies are evaluated at the equilibrium configuration $\mathbf{Q}=0$, suitable for the crude adiabatic representation. In the nuclear Hamiltonian, $\hat{P}_\alpha$ and $\hat{Q}_\alpha$ denote the mass-weighted momentum and coordinate operators of phonon mode $\alpha$, respectively. The corresponding phonon frequencies $\omega_\alpha$ are obtained by diagonalizing the dynamical matrix constructed from the Stillinger--Weber force field~\cite{zhou2013stillinger}. The exciton--nuclear interaction is retained to first order in the phonon coordinates, with coupling constants $V_{nm}^\alpha$ generated from the semiempirical pseudopotential method~\cite{jasrasaria2021interplay}. Higher-order couplings can also be included as discussed in our recent work on the photoluminescence of NCs~\cite{peng2026photoluminescence}.

Fig.~\ref{fig:Fig1_level_spec} panels (a) and (b) compare the bare exciton energies $E_n^0$ and the corresponding polaron-shifted energies $\mathcal{E}_n = E_n^0-\lambda_{nn}$ with reorganization energy $\lambda_{nm}=\sum_\alpha(V_{nm}^\alpha)^2/(2\omega_\alpha^2)$ for the lowest $N_\mathrm{ex}=9$ exciton states near the band edge for both types of NCs.  The energy gaps between neighboring bare excitonic states range from roughly $1$ to $40$~meV. The bare-core CdSe NC exhibits a larger reorganization energy than the CdSe/CdS core--shell structure, leading to more significant polaronic energy shifts. 

The spectral densities for the bare core CdSe NC shown in Fig.~\ref{fig:Fig1_level_spec} panel (c) show significant  couplings to low-frequency phonons below roughly $25$~meV, with contributions from surface torsional and breathing modes~\cite{lin2023theory,hou2025unraveling}. By contrast, for the CdSe/CdS core--shell NC, the coupling is predominantly to higher-frequency optical phonons, with the dominant contribution arising from core vibrations, as shown in Fig.~\ref{fig:Fig1_level_spec} panel (d). In both systems, the spectral densities are governed by the hole channel. The reduced coupling to low-frequency phonons in the core--shell structure is consistent with stronger confinement of the hole to the interior of the NC, which suppresses the interaction with shell surface phonons. The parametrization therefore shows that CdSe core and CdSe/CdS core--shell NCs fall into distinct vibronic regimes: the bare-core system is more strongly coupled to soft lattice modes and has a larger reorganization energy, whereas the core--shell system couples more strongly to optical phonons and has a smaller overall reorganization energy. Both systems, however, exhibit structured spectral densities, with characteristic phonon energies comparable to the gaps between neighboring excitonic states. This places the exciton dynamics in an intermediate-coupling regime, where the comparable electronic and vibrational energy scales render the real-time relaxation dynamics challenging.

\subsection{ML-MCTDH Benchmark}
A numerically exact route to solving the dynamics generated by the model Hamiltonian $\hat{H}$ is to propagate the time-dependent Schr\"odinger equation for the many-body state $\ket{\Psi(t)}$, which lives in the combined excitonic and nuclear Hilbert space. The multi-configuration time-dependent Hartree (MCTDH) approach~\cite{Meyer1990,Manthe1992,Beck2000}, 
together with its multi-layer extension (ML-MCTDH)~\cite{Wang2003,Manthe2008,Vendrell2011,Wang2015},  provides an efficient representation of this wavefunction in terms of time-dependent basis functions for each degree of freedom,
\begin{equation}
    \ket{\Psi(t)} =
    \sum_{n=1}^{N_\mathrm{ex}}
    \sum_{v_1=1}^{n_1}\cdots
    \sum_{v_f=1}^{n_f}
    A_{n,v_1\ldots v_f}(t)\,
    \ket{\psi_n}
    \prod_{\kappa=1}^{f}
    \ket{\varphi_{v_\kappa}^{(\kappa)}(t)},
\end{equation}
where $A_{n,v_1\ldots v_f}(t)$ is the coefficient tensor and $\ket{\varphi_{v_\kappa}^{(\kappa)}(t)}$ denotes the single-particle function (SPF) associated with the $\kappa$th nuclear degree of freedom. In ML-MCTDH, these SPFs are expanded recursively in terms of lower-layer SPFs arranged on a tree tensor network, which greatly alleviates the exponential scaling of the full wavefunction representation. The equations of motion for both the coefficient tensor and the SPFs are obtained variationally from the Dirac-Frenkel principle~\cite{Beck2000}. For finite temperature, we employ the thermofield formulation of ML-MCTDH~\cite{Fischer2021,Borrelli2021}, in which the thermal density operator is mapped onto a pure state in an enlarged Hilbert space, thereby enabling numerically exact benchmark dynamics for exciton cooling at $300$~K.  The ML-MCTDH calculations were
carried out using the Heidelberg MCTDH package~\cite{mctdh:MLpackage}. See {Supporting Information} for more detailed discussion. 

\subsection{Perturbative Quantum Master Equation}
Two classes of approximate methods are also examined. The first is based on a perturbative treatment of the memory kernel within the QME formulation~\cite{nakajima1958quantum,zwanzig1960ensemble,shibata1977generalized}. Here, we work in the polaron-transformed framework~\cite{silbey_variational_1984,jang_theory_2008,nazir_correlation_2009,xu2016non}, where the total Hamiltonian is transformed into 
\begin{equation}
\begin{split}
    \tilde{H} &=e^{\hat{S}} \hat{H} e^{-\hat{S}}=\tilde{H}_\mathrm{ex} + \hat{H}_\mathrm{nu} + \tilde{H}_\mathrm{ex-nu}\\
    \tilde{H}_\mathrm{ex}&= \sum_{n=1}^{N_\mathrm{ex}}\mathcal{E}_n\ket{\psi_n}\bra{\psi_n} \\
     \tilde{H}_\mathrm{ex-nu}&=\sum_{n\neq m}^{N_\mathrm{ex}} \ket{\psi_n}\bra{\psi_m} e^{\hat{S}_n} \sum_{\alpha=1}^{N_\mathrm{nu}} V_{nm}^\alpha \hat{Q}_\alpha e^{-\hat{S}_m}\\
\end{split}
\end{equation}
with the unitary transformation 
\begin{equation}
\begin{split}
    e^{\hat{S}}=\exp\left(-\frac{i}{\hbar}\sum_{n=1}^{N_\mathrm{ex}} \ket{\psi_n}\bra{\psi_n}\hat{S}_n\right),\qquad\hat{S}_n =\sum_{\alpha=1}^{N_\mathrm{nu}} \frac{V_{nn}^\alpha}{\omega_\alpha^2}\hat{P}_\alpha.
\end{split}
\end{equation}
Since the transformation is diagonal in the excitonic basis, the exciton population projector $\ket{\psi_n}\bra{\psi_n}$ is invariant under the polaron transform. Therefore, exciton populations can be obtained directly from the diagonal elements of the reduced density matrix in the polaron-transformed frame $\hat{\rho}_\mathrm{ex}(t)$. The reduced exciton density matrix evolves according to:\cite{zwanzig1960ensemble}
\begin{equation}\label{eq:GME}
\frac{\mathrm{d}}{\mathrm{d}t}\hat{\rho}_\mathrm{ex}(t)
=
-\frac{i}{\hbar}\mathcal{L}_\mathrm{ex}\hat{\rho}_\mathrm{ex}(t)
+\frac{1}{\hbar^2}\int_0^t \mathrm{d}\tau \,\mathcal{K}(t-\tau)\,\hat{\rho}_\mathrm{ex}(\tau),
\end{equation}
where
$\mathcal{L}_\mathrm{ex}\hat{\rho}_\mathrm{ex}
=
\left[\tilde{H}_\mathrm{ex}+\langle \tilde{H}_\mathrm{ex-nu}\rangle,\hat{\rho}_\mathrm{ex}\right]$,
and $\mathcal{K}(t)$ is the memory kernel. In practice, propagating the QME with an exact memory kernel is computationally demanding, so a perturbative treatment of $\mathcal{K}(t)$ is employed~\cite{jang2020dynamics,lai2021simulating}. The choice of the zeroth-order Hamiltonian and interaction term is therefore crucial. Here, following previous work~\cite{peng2024polariton,hou2025unraveling}, we use the diagonal portion of the polaron-transformed Hamiltonian in the diabatic basis $\{\ket{\psi_n}\}$ as the zeroth-order Hamiltonian ($\hat{H}^{(0)}$), and treat the off-diagonal portion ($\hat{H}^{(\mathrm{I})}$) perturbatively: 
\begin{equation}
        \hat{H}^{(0)} = \tilde{H}_\mathrm{ex} + \hat{H}_\mathrm{nu} \qquad \hat{H}^{(\mathrm{I})} =  \tilde{H}_\mathrm{ex-nu}.
\end{equation}
This partitioning absorbs the exciton--phonon couplings that are diagonal in the original exciton basis into the state-specific phonon displacements of the polaron transformation. In the transformed Hamiltonian, their influence enters through the displacement operators in the off-diagonal exciton-transfer terms, and the interaction $\hat{H}^{(\mathrm{I})}$ is treated perturbatively to second order:
\begin{equation}
    \begin{split}
        &\mathcal{K}_{nm,kl}^{(2)}(t) = \left<\delta \hat{H}^{(\mathrm{I})}_{lm}(t) \delta \hat{H}^{(\mathrm{I})}_{nk}\right> + \left<\delta \hat{H}^{(\mathrm{I})}_{kn}(t) \delta \hat{H}^{(\mathrm{I})}_{ml}\right>^{*} \\
        &-\delta_{lm}\sum_r \left<\delta \hat{H}^{(\mathrm{I})}_{nr}(t) \delta \hat{H}^{(\mathrm{I})}_{rk}\right> - \delta_{nk}\sum_r \left<\delta \hat{H}^{(\mathrm{I})}_{mr}(t) \delta \hat{H}^{(\mathrm{I})}_{rl}\right>^{*}
    \end{split}
\end{equation}
where
$\delta\hat{H}^{(\mathrm{I})}
=
\hat{H}^{(\mathrm{I})}-\langle \hat{H}^{(\mathrm{I})}\rangle$, the interaction-picture operator $\delta\hat{H}^{(\mathrm{I})}(t)=e^{\frac{i}{\hbar}\hat{H}^{(0)}t}\delta\hat{H}^{(\mathrm{I})}e^{-\frac{i}{\hbar}\hat{H}^{(0)}t}$ 
and the average is taken with respect to the thermal density matrix
\begin{equation}\label{eq:thermal-quantum}
    \langle \hat{A}\rangle
\equiv
\mathrm{Tr}_{\mathrm{nu}}\!\left\{\hat{A}\hat{\rho}_\mathrm{nu}\right\}, \qquad
\hat{\rho}_\mathrm{nu}
=
\frac{e^{-\beta \hat{H}_\mathrm{nu}}}{\mathrm{Tr}\!\left\{e^{-\beta \hat{H}_\mathrm{nu}}\right\}}.
\end{equation}
The detailed derivation of the memory kernel is provided in the {Supporting Information}.

The presence of strongly coupled low-frequency modes may contribute to the breakdown of the perturbative QME, since these modes can introduce long-lived bath memory and fluctuations outside the perturbative regime. To partially account for this effect, we also consider a frozen-mode QME (FM-QME)~\cite{berkelbach2012reduced,montoya2015extending,teh2019frozen}, in which the low-frequency nuclear vibrations are treated explicitly as static disorder, while the remaining fast modes are treated perturbatively within the same polaron-transformed QME framework. Further details of this formulation are given in the Supporting Information.

\subsection{Mixed Quantum Classical Methods}
Another class of approximate methods is MQC approaches, which treat the nuclear DOFs classically~\cite{Tully1998,stock2005classical,Kapral2006,CrespoOtero2018}. This classical approximation is appropriate for the present systems because the characteristic nuclear frequencies are on the order of, or smaller than, the thermal energy at $300$~K. The starting point is the semiclassical version of the model Hamiltonian, where the nuclear DOFs ($\mathbf{Q}$ and $\mathbf{P}$) are treated as classical dynamical variables:
\begin{equation}
    \hat{H}(\mathbf{P},\mathbf{Q}) = \sum_{\alpha=1}^{N_\mathrm{nu}} \frac{{P}_\alpha^2}{2} + \hat{U}(\mathbf{Q}).
\end{equation}
In  the above, the excitonic potential energy operator is given by:
\begin{equation}
    \begin{split}
        \hat{U}(\mathbf{Q}) &= \sum_{\alpha=1}^{N_\mathrm{nu}}\frac{1}{2}\omega_\alpha^2{Q}_\alpha^2\\ 
    &+ \sum_{n,m=1}^{N_\mathrm{ex}}\left(E_n^0\delta_{nm}+\sum_{\alpha=1}^{N_\mathrm{nu}}V_{nm}^\alpha {Q}_\alpha\right)\ket{\psi_n}\bra{\psi_m}
    \end{split}
\end{equation}
and can also be written in the adiabatic basis $\left\{\ket{\phi_a(\mathbf{Q})}\right\}$
\begin{equation}
    \hat{U}(\mathbf{Q}) = \sum_{a=1}^{N_\mathrm{ex}} U_a(\mathbf{Q})\ket{\phi_a(\mathbf{Q})}\bra{\phi_a(\mathbf{Q})},
\end{equation}
with $\ket{\phi_a(\mathbf{Q})}$ being the eigenstate of $\hat{U}(\mathbf{Q})$ and $U_a(\mathbf{Q})$ being the adiabatic potential energy surface (PES). In MQC dynamics, the time evolution of the excitonic wavefunction $\ket{\psi_\mathrm{ex}(t)}$ follows the time-dependent Schr\"odinger equation, while the nuclear DOFs follow classical equations of motion of the form
\begin{equation}
    \begin{split}
        \dot{Q}_\alpha(t) &= P_\alpha(t)\\
        \dot{P}_\alpha(t) &= F_\alpha(\mathbf{Q}(t))
    \end{split}
\end{equation}
where $F_\alpha(\mathbf{Q})$ is the nuclear force, which depends on the approximation used to incorporate the electronic back-reaction on the nuclei.

Different MQC approaches use different strategies for approximating the nuclear force. In mean-field Ehrenfest dynamics, the nuclei evolve on an effective potential corresponding to a coherent average over electronic states~\cite{Tully1998}
\begin{equation}
    F_\alpha^\mathrm{MF}(\mathbf Q)=-\bra{\psi_\mathrm{ex}(t)}\nabla_\alpha \hat{U}(\mathbf Q)\ket{\psi_\mathrm{ex}(t)}.
\end{equation}
A related family of approaches is provided by semiclassical mapping methods, in which the discrete excitonic states are mapped onto continuous phase-space variables~\cite{MeyerMiller1979,StockThoss1997,Li2013}. Here, we consider spin linearized semiclassical (spin-LSC) approach, which is based on the spin-mapping formalism~\cite{Runeson2019, Runeson2020}. For surface hopping approaches, the nuclear force is evaluated on a single adiabatic surface, referred to as the active surface. If $a(t)$ labels the active surface at time $t$, then the force can be written as
\begin{equation}
    F_\alpha^\mathrm{SH}(\mathbf Q)=-\bra{\phi_{a(t)}(\mathbf{Q})}\nabla_\alpha \hat{U}(\mathbf Q)\ket{\phi_{a(t)}(\mathbf{Q})}.
\end{equation}
Traditional fewest-switches surface hopping (FSSH) chooses the active surface $a(t)$ stochastically based on the electronic populations and derivative couplings~\cite{Tully1990,Subotnik2016}. The recently developed mapping approach to surface hopping (MASH)~\cite{Mannouch2023} combines the mapping formalism with surface-hopping dynamics. It has several multi-state formulations, including the largest-population and size-consistent variants~\cite{Runeson2023,Lawrence2024SizeConsistent}. Here, we use the largest-population prescription, which is simpler and has shown high accuracy for exciton benchmark systems such as the Fenna--Matthews--Olson (FMO) complex~\cite{Runeson2023}. In this approach, the active surface is chosen deterministically as the adiabatic state with the largest electronic population:
\begin{equation}
    a(t) = \operatorname*{argmax}_{b}\left|c_b(t)\right|^2.
\end{equation}
MASH has been shown to outperform FSSH in several model systems due to its more accurate treatment of coherence and its recovery of the quantum--classical equilibrium distribution~\cite{Runeson2023,Mannouch2023,Amati2023,Richardson2025}.

\subsection{Initial Conditions and Population}
Unless otherwise stated, the dynamics is initialized in one of the diabatic excitonic states while the nuclear degrees of freedom remain equilibrated in the uncoupled bath. In the quantum nuclear case, this initial condition is written as
\begin{equation}
    \hat{\rho}^{(\mathrm{d})}(0) = \ket{\psi_n}\bra{\psi_n}\otimes\hat{\rho}_\mathrm{nu}.
\end{equation}
This diabatic preparation corresponds to a vertical Franck--Condon type excitation from the ground state, which creates an exciton at the ground-state equilibrium configuration. In the MQC-based methods, the nuclear ensemble is generated by averaging over trajectories sampled from the classical thermal distribution
\begin{equation}
    {\rho}_\mathrm{nu}(\mathbf{Q}_0, \mathbf{P}_0)
    =
    \frac{e^{-\beta H_\mathrm{nu}(\mathbf{Q}_0, \mathbf{P}_0)}}
    {\frac{1}{{(2\pi\hbar)^{N_\mathrm{nu}}}}\int{\mathrm{d}\mathbf{Q}\mathrm{d}\mathbf{P}}
    e^{-\beta H_\mathrm{nu}(\mathbf{Q}, \mathbf{P})}}.
\label{eq:thermal-classical}
\end{equation}
One can also initialize the system in an adiabatic state, where the exciton is prepared in an eigenstate of the potential operator $\hat{U}(\mathbf{Q})$ at each nuclear geometry:
\begin{equation}
    \hat{\rho}^{(\mathrm{a})}(0)
    =
    \int \mathrm{d}\mathbf{Q}\,
    \rho_{\mathrm{nu}}(\mathbf{Q})\,
    \ket{\phi_a(\mathbf{Q})}\bra{\phi_a(\mathbf{Q})}
    \otimes
    \ket{\mathbf{Q}}\bra{\mathbf{Q}}.
\end{equation}
As before, the nuclear bath is sampled from the same uncoupled equilibrium distribution, cf. Eqs.~\eqref{eq:thermal-quantum} and \eqref{eq:thermal-classical}.

The diabatic and adiabatic populations are defined from the full density operator $\hat{\rho}(t)$ as
\begin{align}
    \mathcal{P}_n^{(\mathrm d)}(t)
    &= \mathrm{Tr}\!\left\{\hat{\rho}(t)\,
    \ket{\psi_n}\bra{\psi_n}\right\},\\
    \mathcal{P}_a^{(\mathrm a)}(t)
    &= \mathrm{Tr}\!\left\{\hat{\rho}(t)\,
    \hat{\Pi}_a^{(\mathrm a)}\right\},
\end{align}
with the adiabatic projector
\begin{equation}
    \hat{\Pi}_a^{(\mathrm a)}
    = \int \mathrm{d}\mathbf{Q}\, \ket{\phi_a(\mathbf{Q})}\bra{\phi_a(\mathbf{Q})} \otimes \ket{\mathbf{Q}}\bra{\mathbf{Q}}.
\end{equation}
The diabatic populations measure the distribution of the excitons defined at the equilibrium geometry, whereas the adiabatic populations track occupation at the instantaneous eigenstate of $\hat{U}(\mathbf{Q})$. The population estimators used for each MQC method are given explicitly in Sec.~V of the Supporting Information.

\section{Results and Discussion}
\subsection{Benchmarking} 
\begin{figure*}[ht!]
    \centering
    \includegraphics[width=0.9\linewidth]{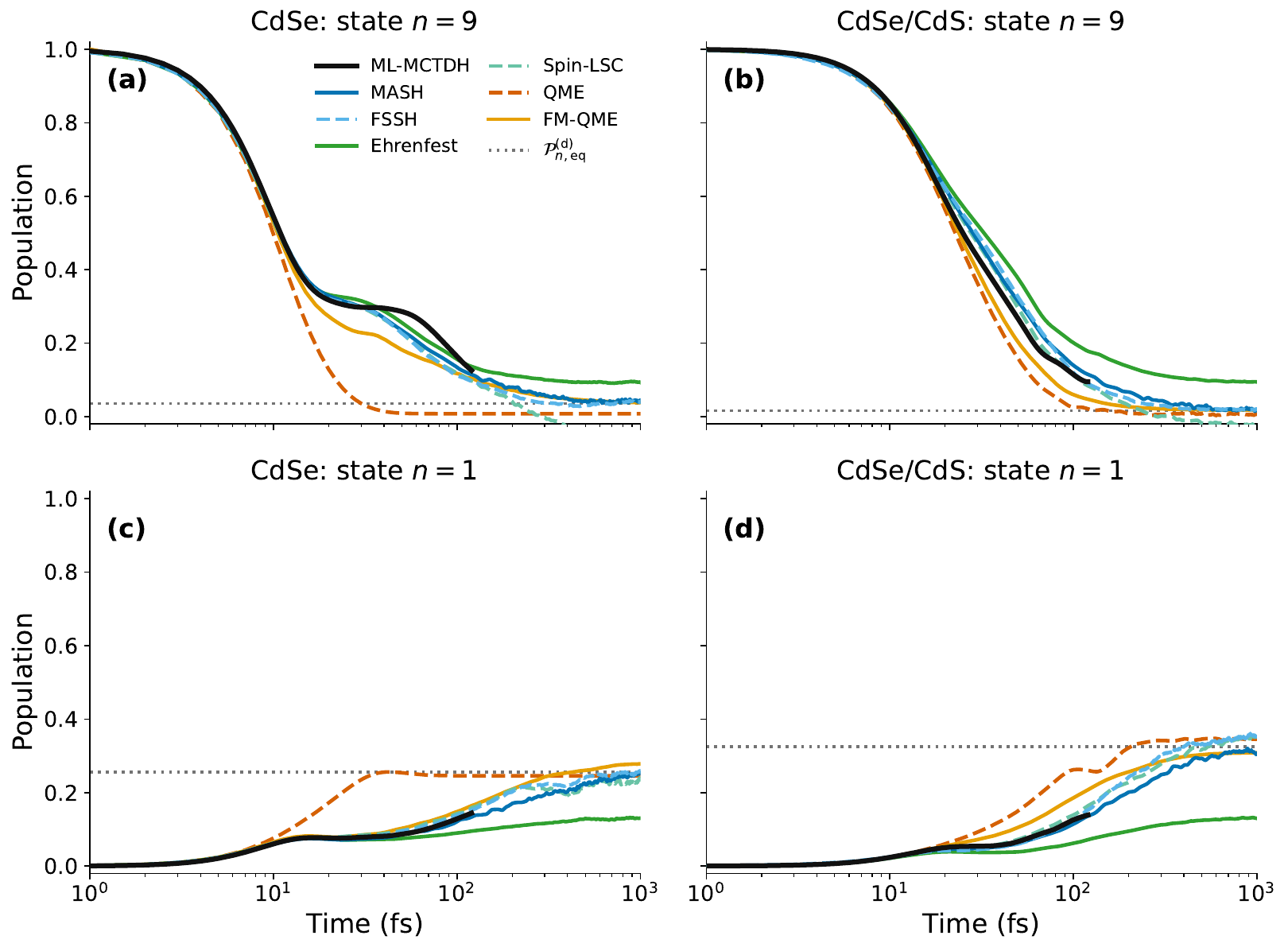}
    \caption{Real-time diabatic population dynamics $\mathcal{P}_n^{(\mathrm d)}(t)$ for CdSe core and CdSe/CdS core--shell NCs for the highest $n=9$ and lowest $n=1$ excitonic states. Thermofield ML-MCTDH method is used as the benchmark. Three classes of approximate methods are compared: perturbative QME, mean-field (Ehrenfest, spin-LSC), and surface hopping (MASH and FSSH) approaches. The black dotted line is the diabatic equilibrium population calculated with PIMC. The benchmark model includes a total of $9$ excitonic states and $360$ modes simulated at $300$~K. }
    \label{fig:Fig2_pop_benchmark}
\end{figure*}
We first examine the diabatic population dynamics $\mathcal{P}_n^{(\mathrm d)}(t)$ at $300$~K. For the model system described above, we obtained converged thermofield ML-MCTDH results for up to $N_\mathrm{ex}=9$ excitonic states and $N_\mathrm{nu}=360$ nuclear modes. In the panels of Fig.~\ref{fig:Fig2_pop_benchmark}, we show the population dynamics of the highest and lowest excitonic states for both CdSe core and CdSe/CdS core--shell NCs starting in the highest excitonic state ($n=9$). We find that exciton cooling is not governed by a single timescale. In particular, for the CdSe core NC, the population of $n=9$ exhibits an ultrafast initial drop within the first $\sim 10$~fs, followed by a slower relaxation over the subsequent few hundred femtoseconds. At the same time, the lowest exciton state ($n=1$), which lies about $150$~meV below the initially populated highest exciton state ($n=9$) at equilibrium geometry, gains population on this short timescale. Since the highest-frequency phonon mode in the model has a vibrational period exceeding $100$~fs, this rapid increase in the population of the lowest excitonic state must arise from fast mixing of exciton states rather than induced by nuclear motion. For the CdSe/CdS core--shell NC, the separation of timescales is less pronounced in the decay of the initially prepared state.

Among the approximate methods tested here, the QME reproduces the initial fast decay most closely, but misses the slower long-time cooling for CdSe core NC and as a result,  overestimates the overall relaxation rate. Treating selected modes as static disorder, as in FM-QME, recovers the separation of timescales, although the resulting dynamics are affected by how the spectral density is partitioned. By contrast, the mixed quantum--classical approaches generally recover the two timescales at least qualitatively, although their quantitative accuracy depends on the specific approximation employed, particularly at longer times. Ehrenfest dynamics tends to over-delocalize the populations and equilibrates earlier than the ML-MCTDH reference. Spin-LSC provides better agreement with the reference than Ehrenfest, but still exhibits noticeable differences, including nonphysical negative populations for some states at long time. Among the surface-hopping methods, both MASH and FSSH perform substantially better than the perturbative and mean-field approaches, providing the most consistent agreement with the ML-MCTDH reference over both the short- and intermediate-time regimes considered here.

\begin{figure*}[ht!]
    \centering
    \includegraphics[width=0.9\linewidth]{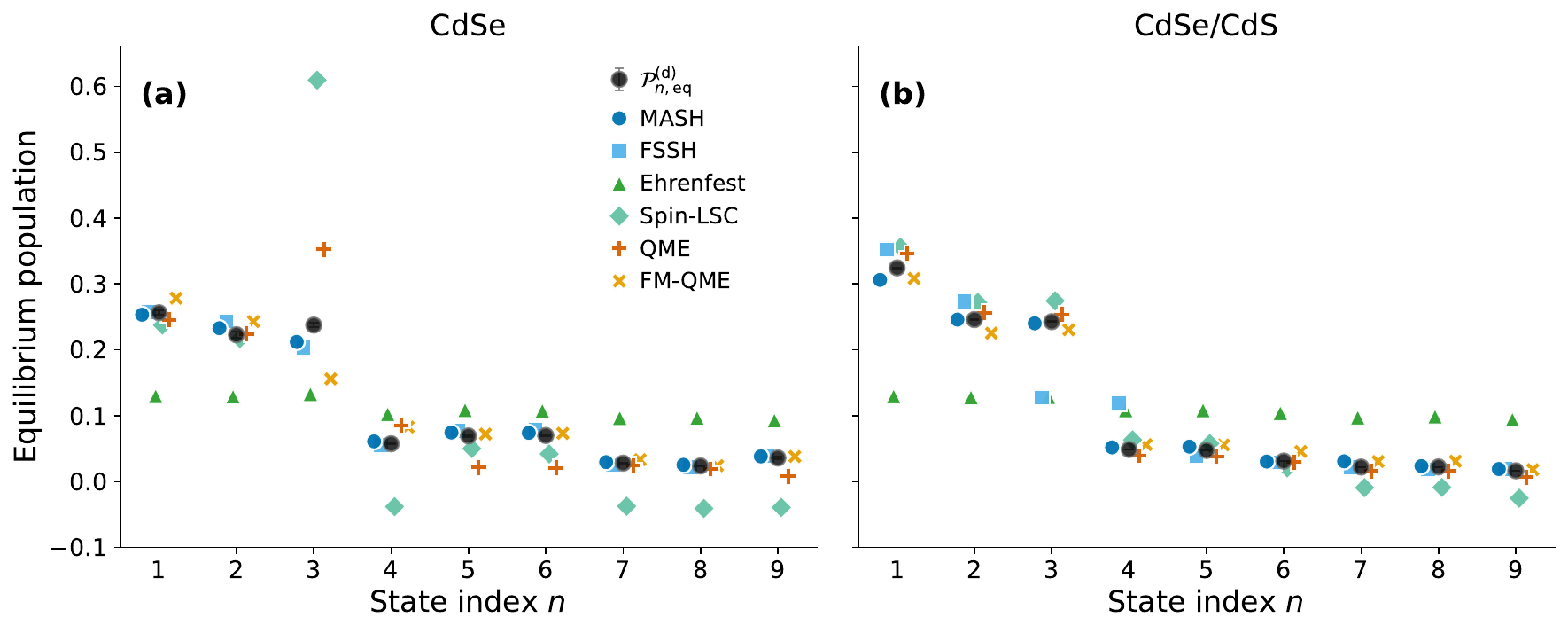}
    \caption{State-resolved diabatic equilibrium populations for (a) CdSe core and (b) CdSe/CdS core--shell NCs at $300$~K. The reference $\mathcal{P}_{n,\mathrm{eq}}^{(\mathrm{d})}$ values are obtained from PIMC (the statistical uncertainty is smaller than the symbol size), using the full exciton--phonon Hamiltonian, while the other symbols show the long-time or method-specific equilibrium populations predicted by MASH, FSSH, Ehrenfest, spin-LSC, QME, and FM-QME. State indices are ordered from the lowest-energy exciton, $n=1$, to the highest state included in the model, $n=9$. }
    \label{fig:Fig3_eq_population}
\end{figure*}
Since ML-MCTDH cannot be efficiently propagated to the thermal equilibrium with a converged basis set for the present models, we instead determine the diabatic equilibrium populations
\begin{equation}
\mathcal{P}^{(\mathrm d)}_{n,\mathrm{eq}}=
\frac{\mathrm{Tr}\left\{
|\psi_n\rangle \langle\psi_n| e^{-\beta \hat H}\right\}}
{\mathrm{Tr} \left\{e^{-\beta \hat H}\right\} }
\label{eq:Peq_d}
\end{equation}
using path-integral Monte Carlo (see the Supporting Information for details)~\cite{ceperley1995path,runeson2025nuclear}.
Fig.~\ref{fig:Fig3_eq_population} compares the PIMC state-resolved diabatic thermal equilibrium populations with the long-time populations predicted by the approximate methods. This comparison is particularly revealing for the CdSe NC, as shown in Fig.~\ref{fig:Fig3_eq_population} panel (a). By construction, the QME approach relaxes to the thermal equilibrium dictated by the zeroth-order Hamiltonian:
\begin{equation}
\mathcal{P}^{(\mathrm d)}_{n,\mathrm{QME}}=\frac{e^{-\beta \mathcal{E}_n}} {\sum_m e^{-\beta \mathcal{E}_m}},
\end{equation}
Given the significant discrepancies between the QME and ML-MCTDH dynamics for the CdSe core NC, one might expect similarly large differences in the diabatic equilibrium populations. Surprisingly, however, the QME approach remains in good overall agreement with the reference, except for a few states with strong off-diagonal couplings $V_{n\ne m}^\alpha$. FM-QME further improves the equilibrium populations and yields results that are even closer to the reference.

\begin{figure*}[ht!]
    \centering
    \includegraphics[width=1.0\linewidth]{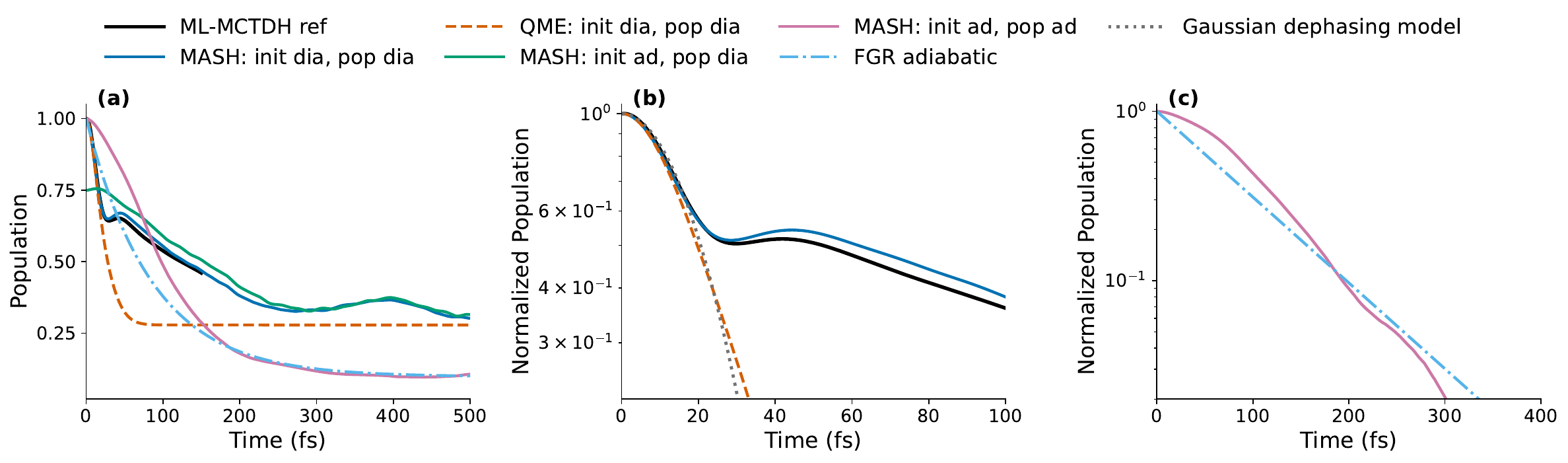}
    \caption{Results for a reduced two-level exciton--phonon model. (a) Population dynamics from ML-MCTDH, MASH, QME and FGR for a chosen initial state and population basis. ``init dia, pop dia'' denotes diabatic initialization and diabatic population measurement; ``init ad, pop dia'' and ``init ad, pop ad'' denote adiabatic initialization with diabatic or adiabatic population measurement, respectively. Diabatic initialization produces an ultrafast initial decay followed by slower relaxation, whereas adiabatic initialization removes the fast component. (b) Short-time normalized population dynamics, compared with the Gaussian dephasing model (cf., Eq.~\eqref{eq:Pdia_short}). (c) Slow relaxation component compared with an adiabatic FGR rate expression.}
    \label{fig:Fig5_tls_timescale}
\end{figure*}
The Ehrenfest approach exhibits a different deficiency, namely over-thermalization across all states, which is a well-known limitation of mean-field dynamics~\cite{parandekar2006detailed}. Spin-LSC also suffers from inaccurate thermal equilibrium and overpopulates state $n=3$ that has a comparatively large reorganization energy. At the same time, it yields unphysical negative equilibrium populations for some higher-energy states, consistent with the zero-point energy leakage problem in mapping approaches~\cite{Cotton2013,Amati2023}. In contrast, both surface-hopping methods recover the equilibrium profile much more faithfully for the core CdSe NC.

For the CdSe/CdS core--shell NC, the equilibrium benchmark is somewhat more forgiving, and all methods except Ehrenfest are nearly quantitative as illustrated in Fig.~\ref{fig:Fig3_eq_population} panel (b). In particular, the perturbative QME agrees well with the exact PIMC. This suggests that the polaron picture is more appropriate for CdSe/CdS core--shell NC, where the excitonic states are dressed predominantly by higher-frequency optical phonons. Somewhat unexpectedly, FSSH deviates from the exact reference for the lowest few states ($n=1-4$), whereas MASH remains in good agreement with the PIMC equilibrium population, as for the previous CdSe core case. As discussed in the {Methods}, MASH is expected to recover the quantum--classical limit of the equilibrium population, which for the present models lies close to the PIMC quantum reference. This is not guaranteed in FSSH and leads to noticeable differences for the equilibrium populations.

Taken together, Figs.~\ref{fig:Fig2_pop_benchmark} and \ref{fig:Fig3_eq_population} show that, among the methods considered here, MASH gives the best overall agreement for both the short- and long-time dynamics and for the thermal equilibrium populations. The success of MASH further suggests that the nuclear motion in the NCs considered here is reasonably well described at the classical limit, provided that the electronic back-reaction is accounted for. In the following, we therefore use MASH as a reliable approximate method to analyze the mechanistic details of exciton relaxation in these NCs.

\subsection{Reduced Two-Level Model}
The distinct timescales observed in the diabatic population dynamics of the bare CdSe NC suggest that exciton relaxation proceeds through a two-step process. To further characterize the two timescales and uncover the cooling mechanism we consider a reduced two-level model, with the two diabatic states chosen to represent the higher-energy initial and lower-lying final exciton states. As shown in Fig.~\ref{fig:Fig5_tls_timescale} panel (a), the diabatic population obtained from both ML-MCTDH (black solid line) and MASH (blue solid line) still exhibits a fast initial decay followed by a slower relaxation, whereas the QME result (orange dashed line) captures only the fast component. Although this reduced model is not intended to quantitatively reproduce the full multistate dynamics of a realistic NC, it provides a minimal framework in which the origin of the two dynamical regimes can be identified more clearly.

For this two-level system, the short-time decay can be expressed analytically in terms of the thermal fluctuations of the off-diagonal diabatic coupling, $V_{12}(\mathbf Q)=\sum_\alpha V_{12}^\alpha Q_\alpha$,
where $1,2$ label the two diabatic states in the reduced model. At short times, the diabatic population can be approximated by
\begin{equation}
\begin{split}
\mathcal{P}_n^{(\mathrm d)}(t)
&\approx
\exp\!\left[
-\frac{t^2 \langle V_{12}(\mathbf Q)^2\rangle}{\hbar^2}
\right] \\
&=
\exp\!\left[
-t^2 \sum_\alpha \frac{(V_{12}^\alpha)^2}{2\hbar\omega_\alpha}
\coth\!\left(\frac{\beta\hbar\omega_\alpha}{2}\right)
\right].
\end{split}
\label{eq:Pdia_short}
\end{equation}
Fig.~\ref{fig:Fig5_tls_timescale} panel (b) compares the dynamics generated by this short-time expression with the normalized population $\tilde{\mathcal P}_n(t) =\frac{\mathcal P_n(t)-\mathcal P_{n,\mathrm{eq}}} {\mathcal P_n(0)-\mathcal P_{n,\mathrm{eq}}}$ and shows that it captures the fast decay observed in both the QME and MASH dynamics quite well. Analyzing Eq.~\eqref{eq:Pdia_short} suggests that the ultrafast component is governed by fluctuations of $V_{12}(\mathbf{Q})$ with the dominant contribution coming from low-frequency modes whose thermal coordinate fluctuations are largest. On this timescale, the nuclei are effectively frozen, so the phonon bath acts as an ensemble of quasi-static realizations of the electronic coupling. The fast Gaussian decay can therefore be viewed as an inhomogeneous dephasing process induced by slow thermal bath fluctuations, rather than as nuclear-motion-driven population transfer. This also explains why the FM-QME works better in Fig.~\ref{fig:Fig2_pop_benchmark}. 


\begin{figure*}[ht!]
    \centering
    \includegraphics[width=0.85\linewidth]{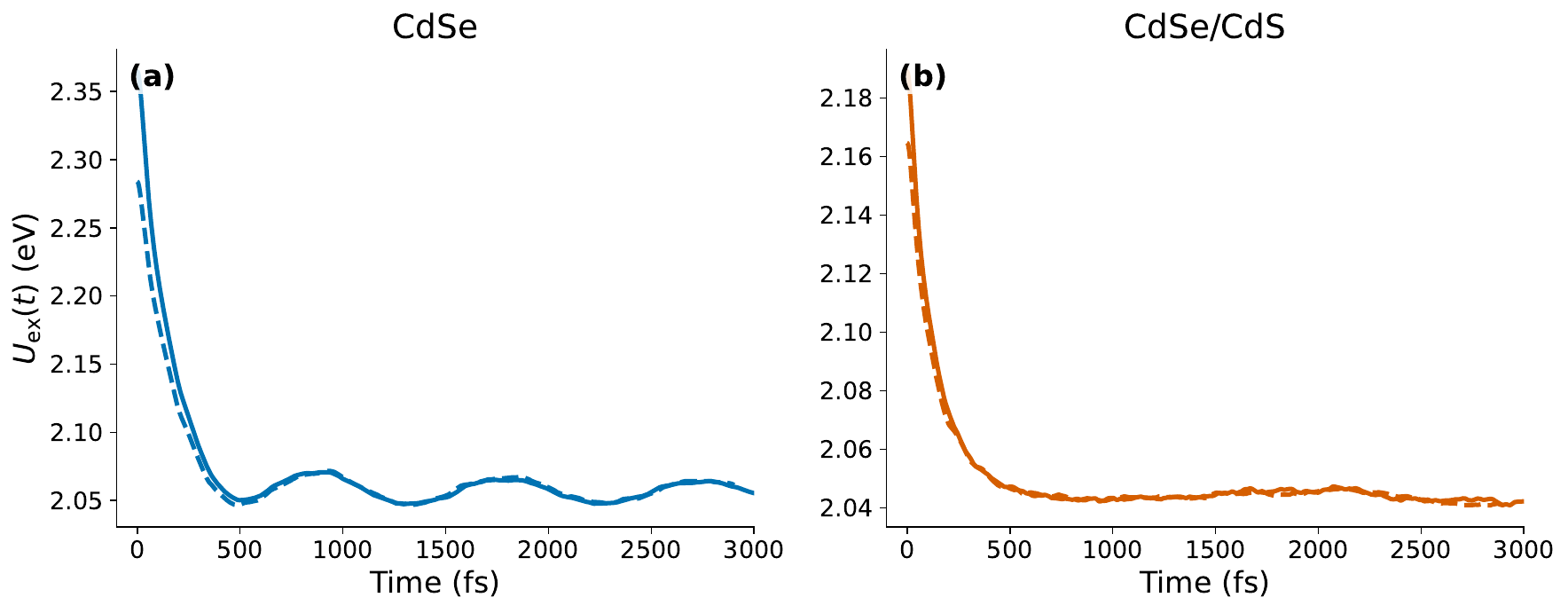}
    \caption{Basis-independent excitonic energy relaxation in (a) CdSe core and (b) CdSe/CdS core--shell NCs. The plotted observable is the averaged excitonic energy $U_{\mathrm{ex}}(t)$ which measures the time evolution of the excitation energy above the band gap. Solid and dashed curves correspond to adiabatic and diabatic initial preparations, respectively. In CdSe core NC, $U_{\mathrm{ex}}(t)$ exhibits a long-time coherent modulation associated with low-frequency surface phonons. }
    \label{fig:Fig6_Uel_Ueq}
\end{figure*}
The appearance of this fast timescale is also closely connected to how the initial state is prepared. Up to this point, we have assumed an initial state given by $\hat{\rho}^{(\mathrm d)}(0)$, corresponding to a vertical Franck--Condon excitation from the ground-state nuclear configuration into a diabatic excited state. When the thermal fluctuations of the off-diagonal coupling, quantified by $\langle V_{12}(\mathbf{Q})^2 \rangle_\mathrm{nu}$, are large, as in the CdSe core NC, the diabatic states undergo rapid mixing, which gives rise to a fast initial change in the diabatic population. By contrast, when the initial state is prepared in an adiabatic state $\ket{\phi_a(\mathbf{Q})}$, this fast component is absent. Indeed, as shown in panel (a) of Fig.~\ref{fig:Fig5_tls_timescale}, MASH dynamics initialized from $\hat{\rho}^{(\mathrm a)}(0)$ exhibits only the slower relaxation.

The slower timescale can therefore be attributed to nuclear-assisted exciton relaxation following the initial diabatic mixing. Because the diabatic excitonic states become substantially mixed after the short induction period, it is more natural to analyze the slower portion of the cooling dynamics using the adiabatic representation. The magenta solid line in panel (a) of Fig.~\ref{fig:Fig5_tls_timescale} shows the MASH adiabatic population $\mathcal{P}_a^{(\mathrm a)}(t)$ for an initial condition prepared in the upper adiabatic state. Unlike the diabatic population, the adiabatic population is governed predominantly by a single relaxation timescale, which agrees well with the slower component of the diabatic population decay.  For this two-level model, we also derive a classical Fermi's golden rule (FGR) rate expression for transitions between the two adiabatic states induced by derivative couplings (see Supporting Information for details), and use it to construct the corresponding population-relaxation dynamics. As shown in panel (c) of Fig.~\ref{fig:Fig5_tls_timescale}, this adiabatic perturbative treatment successfully captures the slower, nuclear-assisted relaxation timescale.

Taken together, the separation between the initial diabatic mixing and the subsequent nuclear-assisted adiabatic relaxation helps explain why and when the QME provides accurate results and when it fails. In the QME, the polaron-dressed off-diagonal couplings are treated perturbatively, so its accuracy is controlled by the magnitude of the thermal fluctuations in $V_{12}(\mathbf Q)$, quantified by $\langle V_{12}(\mathbf Q)^2\rangle$. In the classical limit, these fluctuations satisfy $\langle V_{12}(\mathbf Q)^2\rangle \approx 2k_{\mathrm B}T\lambda_{12}$, and therefore increase with both temperature and the off-diagonal reorganization energy. In CdSe core NCs, low-frequency phonons contribute strongly to $\lambda_{12}$, producing large fluctuations and very short dephasing times. { At $300$~K, these low-frequency modes are thermally populated and can contribute substantially to the coupling, whereas the higher-frequency optical modes that dominate the coupling for the core--shell NC are less populated.} Under these conditions, the QME captures mainly the initial ultrafast decay, but misses the slower relaxation that emerges at later times. In CdSe/CdS core--shell NCs, by contrast, the weaker coupling to low-frequency phonons leads to smaller fluctuations in the off-diagonal coupling, suppresses the ultrafast component, and extends the regime in which the perturbative QME remains reliable.

%


\subsection{Basis-Independent Relaxation}
Although diabatic and adiabatic populations provide a useful microscopic interpretation of the simulated dynamics, ultrafast experiments do not usually probe such basis-specific quantities directly. In colloidal CdSe NCs, exciton relaxation rates are most commonly inferred from transient absorption (TA) or multidimensional electronic spectroscopies, which measure optical response functions rather than state populations themselves~\cite{kambhampati2011hot,ghosh2025atomistically}. Recent work has further emphasized that conventional TA does not cleanly resolve the full hot-exciton cooling pathway, whereas multidimensional spectroscopies can more directly disentangle coherence and population-transfer contributions~\cite{ghosh2025atomistically}. In this context, our calculations suggest that an ultrafast signal appearing within the first few tens of femtoseconds does not necessarily correspond uniquely to nuclear-assisted relaxation, but may also reflect rapid state mixing driven by electronic dephasing, depending on how the initial excited state is prepared. This interpretation is consistent with recent measurements on CdSe-based NCs that reported a cooling signature on the order of $\sim 30$~fs~\cite{brosseau2023ultrafast,ghosh2025atomistically}.

To connect the simulated dynamics more directly to experimentally relevant measures of energy relaxation, we next consider the basis-independent excitonic energy,
\begin{equation}
U_{\mathrm{ex}}(t)=\int d\mathbf P\, d\mathbf Q \,
\mathrm{Tr}_\mathrm{ex}
\left\{
\hat{U}_\mathrm{ex}(\mathbf Q)\,
\hat{\rho}_{\mathrm{MASH}}(\mathbf P,\mathbf Q;t)
\right\},
\end{equation}
where the exciton potential operator is defined as
\begin{equation}
\hat{U}_\mathrm{ex}(\mathbf Q)
=
\sum_{n,m=1}^{N_\mathrm{ex}}
\ket{\psi_n}\bra{\psi_m}
\left[
E_n^0\delta_{nm}
+
\sum_{\alpha=1}^{N_{\mathrm{nu}}}V_{nm}^\alpha {Q}_{\alpha}
\right].
\end{equation}
Unlike the diabatic or adiabatic populations, $U_{\mathrm{ex}}(t)$ is independent of the chosen excitonic representation and therefore provides a direct measure of the net excitonic energy during relaxation.

Fig.~\ref{fig:Fig6_Uel_Ueq} panel (a) shows $U_{\mathrm{ex}}(t)$ for the CdSe NC, with initial conditions prepared in the highest-energy excitonic state at the ground-state equilibrium geometry in the chosen representation. The solid line uses the adiabatic preparation, $\hat{\rho}^{(\mathrm a)}(0)$, whereas the dashed line uses the diabatic preparation, $\hat{\rho}^{(\mathrm d)}(0)$. The choice of initial basis primarily affects the initial value $U_{\mathrm{ex}}(0)$, while leaving the overall relaxation timescale largely unchanged. For the CdSe core NC, $U_{\mathrm{ex}}(t)$ decays toward the band-edge energy within approximately $500$~fs and then exhibits oscillations with an amplitude of about $10$~meV at longer times. The corresponding oscillation period is consistent with coherent modulation by low-frequency phonons with energy $\sim 5$~meV, which are associated primarily with surface vibrations~\cite{jasrasaria2021interplay}. By contrast, in the CdSe/CdS core--shell NC, $U_{\mathrm{ex}}(t)$ exhibits a smoother decay with little long-time oscillatory modulation, as shown in Fig.~\ref{fig:Fig6_Uel_Ueq} panel (b). We therefore view $U_{\mathrm{ex}}(t)$ as a more direct proxy for exciton cooling than basis-specific populations, particularly in situations where the early-time dynamics contains both dephasing-like state mixing and slower nuclear-assisted relaxation.

\section{Conclusion}
In this work, we investigated hot-exciton relaxation in atomistically modeled CdSe core and CdSe/CdS core--shell NCs by combining numerically converged thermofield ML-MCTDH benchmark dynamics, PIMC equilibrium calculations, and a range of approximate dynamical methods. Our results show that exciton cooling in these systems is not generically governed by a single relaxation timescale. In particular, the bare CdSe core NC exhibits two clearly separated dynamical regimes: an ultrafast component on the order of $10$~fs and a slower relaxation on the order of $100$~fs, whereas the CdSe/CdS core--shell NC is dominated primarily by the slower component.

Analysis of a reduced two-level model clarifies the physical origin of these two regimes. The ultrafast component arises from rapid diabatic state mixing induced by thermal fluctuations of the off-diagonal exciton--phonon coupling, with low-frequency phonons playing a dominant role through their large thermal fluctuations. This initial decay is therefore better understood as dephasing-like mixing under quasi-static bath disorder rather than as nuclear-assisted population transfer. The slower component, by contrast, reflects genuine nuclear-assisted relaxation between adiabatic excitonic states. This separation between short-time diabatic mixing and longer-time adiabatic relaxation provides a simple mechanistic framework for interpreting hot-exciton cooling beyond the conventional phonon-bottleneck picture.

These findings also explain the varying performance of approximate dynamical methods. The perturbative QME captures the initial fast decay well, but can fail in the diabatic representation when low-frequency phonons generate large fluctuations in the off-diagonal coupling, as in bare CdSe NCs. At the same time, the analysis of the reduced two-level model shows that the slower relaxation reflects nuclear-assisted transitions between adiabatic states. In contrast, the weaker low-frequency coupling in CdSe/CdS core--shell NCs places the system closer to the perturbative regime already in the diabatic representation, leading to much better QME performance throughout. Among the MQC approaches considered here, MASH provides the most consistent agreement with the benchmark dynamics and with the equilibrium populations, outperforming mean-field and standard mapping approaches and also improving upon FSSH in the equilibrium limit.

Finally, by analyzing the basis-independent excitonic energy, we showed that the overall cooling timescale is less sensitive to the choice of excitonic representation than the state populations themselves, while still revealing important differences between the two materials. In the CdSe core NC, the energy relaxation is accompanied by long-time oscillatory modulation from low-frequency surface phonons, whereas the CdSe/CdS core--shell NC displays a smoother decay. This suggests that experimentally observed sub-picosecond cooling signals may contain contributions from both rapid electronic dephasing and slower nuclear-assisted relaxation, and that care is needed when assigning early-time spectroscopic signatures to a single microscopic mechanism. 

\section{Supporting Information}
The Supporting Information includes details of the model Hamiltonian construction and exciton--phonon parameterization, thermofield ML-MCTDH convergence, derivations of the QME and FM-QME formulations, MQC population estimators, PIMC equilibrium calculations, and the reduced two-level-model analysis.

\begin{acknowledgments}
We would like to thank Professor David T. Limmer and Dr. Eric R. Heller for fruitful discussions. This work was supported by the National Science Foundation Division of Chemistry, under the Chemical Theory, Models and Computational Methods (CTMC) program, grant number CHE-2449564 and the U.S. National Science Foundation Science and Technology Center (STC) for Integration of Modern Optoelectronic Materials on Demand (IMOD) under Cooperative Agreement No. DMR-2019444.  This research used resources of the National Energy Research Scientific Computing Center, a DOE Office of Science User Facility supported by the Office of Science of the U.S. Department of Energy under Contract No. DE-AC02-05CH11231 using NERSC award BES-ERCAP0032503. ER would like to thank the Israel Science Foundation for support (No. 4085/25). 
\end{acknowledgments}

\subsection*{Conflict of Interest Statement}
The authors have no conflicts to disclose.

\section*{Data Availability Statement}
The data that support the findings of this study are available
from the corresponding authors upon reasonable request.
\bibliography{main}
\end{document}


\title{Supporting Information: Full Quantum and Mixed Quantum--Classical Dynamics of Hot Exciton Cooling in Quantum Dots}

\author{Bokang Hou}
\affiliation{Department of Chemistry, University of California, Berkeley, California, 94720, United States}
\author{Johan E. Runeson}
\affiliation {Institute of Physics, University of Freiburg, Hermann-Herder-Straße 3, 79104 Freiburg, Germany;}
\author{Samuel L. Rudge}
\affiliation {Institute of Physics, University of Freiburg, Hermann-Herder-Straße 3, 79104 Freiburg, Germany;}
\author{Salvatore Gatto}
\affiliation {Institute of Physics, University of Freiburg, Hermann-Herder-Straße 3, 79104 Freiburg, Germany;}
\author{Hans-Dieter Meyer}
\affiliation {Theoretische Chemie, Physikalisch-Chemisches Institut, Universität Heidelberg, INF 229, D-69120 Heidelberg, Germany;}
\author{Michael Thoss}
\affiliation {Institute of Physics, University of Freiburg, Hermann-Herder-Straße 3, 79104 Freiburg, Germany;}
\author{Eran Rabani}
\affiliation{Fritz Haber Center for Molecular Dynamics, The Institute of Chemistry and The Institute of Applied Physics, The Hebrew University of Jerusalem, Jerusalem 91904, Israel}
\affiliation{Department of Chemistry, University of California, Berkeley, California, 94720, United States}
\affiliation{Materials Sciences Division, Lawrence Berkeley National Laboratory, Berkeley, California, 94720, United States}
\maketitle



\section{Nanocrystal configuration}
The theoretical calculations were performed for two nanocrystal (NC) structures: a bare CdSe core NC and a CdSe/CdS core-shell NC. Both structures were generated from a $3$~nm CdSe core cut from bulk wurtzite CdSe, which contains 222 Cd atoms and 222 Se atoms. The core-shell structure was then obtained by adding a $2$-monolayer (ML) CdS shell around the same CdSe core, giving a total composition of 753 Cd atoms, 252 Se atoms, and 501 S atoms.

For both the bare-core and core-shell NCs, the Stillinger--Weber force field~\cite{zhou2013stillinger} was utilized to determine the equilibrium structure and normal vibrational modes using LAMMPS~\cite{plimpton1995fast}. Following minimization, the outermost layer was represented by ligand-passivation pseudopotentials~\cite{rabani1999electronic}.

\section{Nanocrystal excited state calculation}

\subsection{Semi-empirical pseudopotential and filter diagonalization}

The electronic structure calculations were performed within the semi-empirical pseudopotential method for CdSe and CdSe/CdS NCs.\cite{rabani1999electronic,jasrasaria2022simulations} We used the same local, screened pseudopotential functions as described in ref.~\citenum{jasrasaria2022simulations} for Cd, Se, and S. These pseudopotentials were fitted to reproduce the bulk band structures, effective masses, and deformation potentials of CdSe and CdS. For an atom of type \(\mu\), the local pseudopotential in momentum space is
\begin{equation}
\tilde v_\mu(q)
=
a_0^\mu
\left[
1+a_4^\mu\operatorname{Tr}\bm{\epsilon}_\mu
+a_5^\mu\left(\operatorname{Tr}\bm{\epsilon}_\mu\right)^3
\right]
\frac{q^2-a_1^\mu}{a_2^\mu\exp\left(a_3^\mu q^2\right)-1},
\label{eq:si_pseudopotential}
\end{equation}
where \(\bm{\epsilon}_\mu\) is the local strain associated with atom type \(\mu\), evaluated from the nearest-neighbor tetrahedron. The pseudopotential parameters used here are listed in Table~\ref{tab:si_pseudopotential_parameters}.

\begin{table}[h]
\caption{Semi-empirical pseudopotential parameters for Cd, Se, and S. All values are given in atomic units.}
\label{tab:si_pseudopotential_parameters}
\begin{tabular}{lrrrrrr}
\hline\hline
Element & \(a_0\) & \(a_1\) & \(a_2\) & \(a_3\) & \(a_4\) & \(a_5\) \\
\hline
Cd & \(-31.4518\) & \(1.3890\) & \(-0.0502\) & \(1.6603\) & \(0.0586\) & \(0\) \\
Se & \(8.4921\) & \(4.3513\) & \(1.3600\) & \(0.3227\) & \(0.1746\) & \(-33\) \\
S  & \(7.6697\) & \(4.5192\) & \(1.3456\) & \(0.3035\) & \(0.2087\) & \(0\) \\
\hline\hline
\end{tabular}
\end{table}

The single-particle electron and hole states near the bottom of the conduction band and the top of the valence band were generated on real-space grids with spacings smaller than \(0.8\)~a.u. We used the filter-diagonalization technique to target states near the lowest unoccupied molecular orbital (LUMO) and highest occupied molecular orbital (HOMO) energies, respectively.~\cite{toledo2002very} The resulting quasiparticle states are eigenstates of the single-particle pseudopotential Hamiltonian and provide the electron and hole basis for the excitonic calculation.

\subsection{Bethe-Salpeter equation}


The exciton states were constructed as linear combinations of electron-hole product states with coefficients obtained by solving the Bethe-Salpeter equation (BSE) within the static screening approximation.~\cite{rohlfing2000electron} The $n$-th exciton wavefunction can be written as:
\begin{equation}\label{eq:bse}
|\psi_{n}\rangle=\sum_{a\in\mathrm{elec}}\sum_{i\in\mathrm{hole}}c_{ai}^{n}|\varphi_{a}\rangle\otimes|\varphi_{i}\rangle,
\end{equation}
where $\varphi_{a}$ and $\varphi_{i}$ are the $a$-th and $i$-th electron and hole states, respectively. $\left\{ c_{ai}^{n}\right\}$ are the BSE coefficients obtained with a static dielectric constant $\epsilon=5$. Here, we used 60 hole states and 80 electron states to converge the excited state calculations. 

\subsection{Exciton--phonon coupling parameterization}
\label{sec:si_expc_parameterization}

The exciton--phonon coupling was expanded to first order in the displacement from the equilibrium NC geometry, giving the linear vibronic coupling Hamiltonian used throughout this work. We first evaluate exciton-nuclear matrix elements with respect to Cartesian atomic displacements,
\begin{equation}
V_{nm}^{\mu k}
=
\left\langle
\psi_n
\left|
\left.
\frac{\partial \hat{v}_\mu(|\mathbf{r}-\mathbf{R}_\mu|)}{\partial R_{\mu k}}
\right|_{\mathbf R_0}
\right|
\psi_m
\right\rangle,
\label{eq:si_expc_cartesian}
\end{equation}
where \(\mu\) labels atoms, \(k\in\{x,y,z\}\) and \(\mathbf R_0\) is the equilibrium geometry. Using the BSE expansion in Eq.~\eqref{eq:bse}, these matrix elements can be evaluated from the BSE coefficients and the single-particle derivative matrix elements as
\begin{equation}
V_{nm}^{\mu k}
=
\sum_{abi}
c_{ai}^{n*}c_{bi}^{m}
v'_{ab,\mu k}
-
\sum_{aij}
c_{ai}^{n*}c_{aj}^{m}
v'_{ij,\mu k},
\label{eq:si_expc_bse}
\end{equation}
with
\begin{equation}
v'_{rs,\mu k}
=
\int d\mathbf r\,
\varphi_r^*(\mathbf r)
\left.
\frac{\partial \hat v_{\mu}(\mathbf r-\mathbf R_\mu)}
{\partial R_{\mu k}}
\right|_{\mathbf R_0}
\varphi_s(\mathbf r).
\label{eq:si_single_particle_derivative}
\end{equation}
The first term in Eq.~\eqref{eq:si_expc_bse} is the electron contribution, while the second term is the hole contribution.

The Cartesian exciton-nuclear couplings are transformed to normal-mode coordinates using the eigenvectors of the mass-weighted dynamical matrix obtained from the Stillinger--Weber force field,
\begin{equation}
V_{nm}^{\alpha}
=
\sum_{\mu,k}
\frac{1}{\sqrt{M_\mu}}
e_{\alpha,\mu k}V_{nm}^{\mu k},
\label{eq:si_expc_normal_mode}
\end{equation}
where \(e_{\alpha,\mu k}\) is the \(\mu k\) component of normal mode \(\alpha\), \(M_\mu\) is the atomic mass, and \(\omega_\alpha\) is the corresponding phonon frequency.

\subsection{Thermofield ML-MCTDH Dynamics}
\label{sec:si_thermofield}

Finite-temperature ML-MCTDH simulations were performed using the thermofield formulation, which represents the thermal vibrational density matrix as a pure state in an enlarged Hilbert space rather than propagating a density operator in Liouville space.~\cite{Fischer2021} We consider the model linear vibronic Hamiltonian
\begin{equation}
\hat H =
\sum_{n=1}^{N_{\rm ex}} E_n^0 |\psi_n\rangle\langle \psi_n|
+
\sum_{\alpha=1}^{N_{\rm nu}}
\frac{1}{2}
\left(\hat P_\alpha^2+\omega_\alpha^2\hat{Q}_\alpha^2\right)
+
\sum_{n,m=1}^{N_{\rm ex}}
|\psi_n\rangle\langle \psi_m|
\sum_{\alpha=1}^{N_{\rm nu}}
V_{nm}^{\alpha}\hat Q_\alpha ,
\label{eq:si_lvc_hamiltonian}
\end{equation}
where \(\hat Q_\alpha=\sqrt{\dfrac{\hbar}{2\omega_\alpha}}(\hat a_\alpha+\hat a_\alpha^\dagger)\) and
\(\hat P_\alpha=-i\sqrt{\dfrac{\hbar\omega_\alpha}{2}}(\hat a_\alpha-\hat a_\alpha^\dagger)\). The initial density matrix is taken to be factorized between a selected excitonic state and a thermal vibrational bath,
\begin{equation}
\hat \rho(0)
=
|\psi_n\rangle\langle \psi_n|
\otimes
\frac{e^{-\beta \hat H_{\mathrm{nu}}}}{Z_{\mathrm{nu}}},
\qquad
\hat H_{\mathrm{nu}} =
\sum_{\alpha=1}^{N_{\rm nu}}
\hbar\omega_\alpha
\left(\hat a_\alpha^\dagger \hat a_\alpha+\frac{1}{2}\right).
\label{eq:si_initial_density}
\end{equation}

In the thermofield representation, each physical vibrational mode in Hilbert space $\mathcal{H}_\mathrm{nu}$ is paired with an auxiliary, or tilde mode in space $\tilde{\mathcal{H}}_\mathrm{nu}$. The thermal bath density matrix is purified as the thermofield vacuum in space $\mathcal{H}_\beta = \mathcal{H}_\mathrm{nu}\otimes \tilde{\mathcal{H}}_\mathrm{nu}$
\begin{equation}
|\mathbf{0}_\beta\rangle
=
\frac{1}{\sqrt{Z_\mathrm{nu}}}
\sum_{\mathbf n}
e^{-\beta E_{\mathbf n}/2}
|\mathbf n,\tilde{\mathbf n}\rangle ,
\qquad
E_{\mathbf n}
=
\sum_{\alpha=1}^{N_{\rm nu}}
\hbar\omega_\alpha n_\alpha ,
\label{eq:si_thermal_vacuum}
\end{equation}
such that expectation values of physical operators reproduce thermal averages. Equivalently,
\begin{equation}
|\mathbf{0}_\beta\rangle
=
e^{-i\hat G}
|\mathbf{0},\tilde{\mathbf 0}\rangle ,
\qquad
\hat G
=
-i\sum_{\alpha=1}^{N_{\rm nu}}
\theta_\alpha
\left(
\hat a_\alpha \hat{\tilde a}_\alpha
-
\hat a_\alpha^\dagger \hat{\tilde a}_\alpha^\dagger
\right),
\label{eq:si_bogoliubov_generator}
\end{equation}
with the temperature-dependent Bogoliubov angle
\begin{equation}
\theta_\alpha
=
\operatorname{arctanh}
\left(e^{-\beta\hbar\omega_\alpha/2}\right).
\label{eq:si_theta}
\end{equation}
For the present simulations only the vibrational bath is thermally populated, whereas the electronic subsystem is initialized in a pure excitonic state. Therefore, an auxiliary electronic space is not required. The thermofield initial state is
\begin{equation}
|\psi_\beta(0)\rangle
=
|\psi_n\rangle
|\mathbf{0}_\beta\rangle .
\label{eq:si_tf_initial_state}
\end{equation}

Instead of propagating the thermofield state with a temperature-dependent initial vibrational state, we apply the inverse Bogoliubov transformation to the Hamiltonian and propagate the zero-temperature vacuum in the doubled vibrational space. Defining
\begin{equation}
|\Psi_\beta(t)\rangle
=
e^{i\hat G}
|\psi_\beta(t)\rangle ,
\end{equation}
the time evolution is governed by
\begin{equation}
\frac{\partial}{\partial t}
|\Psi_\beta(t)\rangle
=
-\frac{i}{\hbar}
\hat{\bar H}_\theta
|\Psi_\beta(t)\rangle ,
\qquad
\hat{\bar H}_\theta
=
e^{i\hat G}
\hat{\bar H}
e^{-i\hat G},
\label{eq:si_tf_schrodinger}
\end{equation}
where
\begin{equation}
\hat{\bar H}
=
\hat H
-
\hat{\tilde H}_{\mathrm{nu}} .
\end{equation}
The auxiliary copy of the physical nuclear Hamiltonian is denoted as 
\begin{equation}
\hat{\tilde H}_{\mathrm{nu}}
=
\sum_{\alpha=1}^{N_{\rm nu}}
\hbar\omega_\alpha\left(
\hat{\tilde a}_\alpha^\dagger \hat{\tilde a}_\alpha+\frac{1}{2}\right)
\end{equation}
which acts only on the tilde vibrational modes. The minus sign in the auxiliary bath Hamiltonian ensures that the thermofield propagation reproduces the correct finite-temperature dynamics of the physical density matrix. The Bogoliubov transformation gives
\begin{align}
e^{i\hat G}\hat a_\alpha e^{-i\hat G}
&=
\hat a_\alpha \cosh\theta_\alpha
+
\hat{\tilde a}_\alpha^\dagger \sinh\theta_\alpha ,
\\
e^{i\hat G}\hat{\tilde a}_\alpha e^{-i\hat G}
&=
\hat{\tilde a}_\alpha \cosh\theta_\alpha
+
\hat a_\alpha^\dagger \sinh\theta_\alpha .
\end{align}
Using these relations, the transformed Hamiltonian becomes
\begin{align}
\hat{\bar H}_\theta
=&
\sum_{n=1}^{N_{\rm ex}}
E_n^0|\psi_n\rangle\langle \psi_n|
+
\sum_{\alpha=1}^{N_{\rm nu}}
\frac{1}{2}
\left[
\hat P_\alpha^2+\omega_\alpha^2\hat Q_\alpha^2
-
\left(
\hat{\tilde P}_\alpha^2+\omega_\alpha^2\hat{\tilde Q}_\alpha^2
\right)
\right]
\nonumber \\
&+
\sum_{n,m=1}^{N_{\rm ex}}
|\psi_n\rangle\langle \psi_m|
\sum_{\alpha=1}^{N_{\rm nu}}
V_{nm}^{\alpha}
\left(
\hat Q_\alpha \cosh\theta_\alpha
+
\hat{\tilde Q}_\alpha \sinh\theta_\alpha
\right).
\label{eq:si_tf_hamiltonian}
\end{align}
Thus finite temperature enters through the temperature-dependent rescaling of the physical coupling, \(\cosh\theta_\alpha\), and through the coupling to the auxiliary coordinate, \(\sinh\theta_\alpha\). In the transformed representation the initial state is simply
\begin{equation}
|\Psi_\beta(0)\rangle
=
|\psi_e\rangle
|\mathbf{0},\tilde{\mathbf 0}\rangle .
\label{eq:si_tf_initial_transformed}
\end{equation}

The wavefunction \( |\Psi_\beta(t)\rangle \) is then propagated with ML-MCTDH in the enlarged vibrational space containing both physical and auxiliary modes. Schematically, the thermofield ML-MCTDH wavefunction is expanded as
\begin{equation}
    \ket{\Psi_\beta(t)} =
    \sum_{n=1}^{N_\mathrm{ex}}
    \sum_{v_1=1}^{n_1}\cdots
    \sum_{\tilde{v}_f=1}^{\tilde{n}_f}
    A_{n,v_1\ldots v_f,\tilde{v}_1\ldots\tilde{v}_f}(t;\beta)\,
    \ket{\psi_n}
    \prod_{\kappa=1}^{f}
    \ket{\varphi_{v_\kappa}^{(\kappa)}(t;\beta)}\prod_{\tilde{\kappa}=1}^{f}
    \ket{\tilde\varphi_{v_{\tilde{\kappa}}}^{({\tilde{\kappa}})}(t;\beta)},
\end{equation}
where indices label the configurations of time-dependent single-particle functions for both the physical and auxiliary vibrational modes. The equations of motion for the expansion coefficients and single-particle functions are obtained from the Dirac--Frenkel variational principle at fixed temperature,
\begin{equation}
\left\langle
\delta\Psi_\beta(t)
\left|
i\hbar\frac{\partial}{\partial t}
-
\hat{\bar H}_\theta
\right|
\Psi_\beta(t)
\right\rangle
=0.
\label{eq:si_dirac_frenkel_tf}
\end{equation}
For observables acting only on the physical Hilbert space,
\begin{equation}
\langle \hat O(t)\rangle_\beta
=
\langle \Psi_\beta(t)|
\hat O_\theta
|\Psi_\beta(t)\rangle ,
\qquad
\hat O_\theta
=
e^{i\hat G}\hat O e^{-i\hat G}.
\label{eq:si_tf_observable}
\end{equation}
In particular, excitonic population operators commute with the vibrational Bogoliubov transformation, so their expectation values can be evaluated directly from \( |\Psi_\beta(t)\rangle \).

\section{Quantum Master Equation}
The total Hamiltonian in the polaron framework is given by~\cite{silbey_variational_1984,jang_theory_2008,nazir_correlation_2009,xu2016non}
\begin{equation}
\begin{split}
    \Tilde{H} &=e^{\hat{S}} \hat{H} e^{-\hat{S}}=\tilde{H}_\mathrm{ex} + \hat{H}_\mathrm{nu} + \tilde{H}_\mathrm{ex-nu}\\
    \tilde{H}_\mathrm{ex}&= \sum_{n=1}^{N_\mathrm{ex}}\mathcal{E}_n\ket{\psi_n}\bra{\psi_n} \\
     \tilde{H}_\mathrm{ex-nu}&=\sum_{n\neq m}^{N_\mathrm{ex}} \ket{\psi_n}\bra{\psi_m} e^{\hat{S}_n} \sum_{\alpha=1}^{N_\mathrm{nu}} V_{nm}^\alpha \hat{Q}_\alpha e^{-\hat{S}_m}\\
\end{split}
\end{equation}
with the unitary transformation 
\begin{equation}
\begin{split}
    e^{\hat{S}}&=\exp\left(-\frac{i}{\hbar}\sum_{n=1}^{N_\mathrm{ex}} \ket{\psi_n}\bra{\psi_n}\hat{S}_n\right), \hspace{0.25cm} \hat{S}_n =\sum_{\alpha=1}^{N_\mathrm{nu}} \frac{V_{nn}^\alpha}{\omega_\alpha^2}\hat{P}_\alpha.
\end{split}
\end{equation}
The reduced exciton density matrix in the polaron framework, $\hat{\rho}_\mathrm{ex}(t)$, evolves according to
\begin{equation}\label{eq:GME}
\frac{\mathrm{d}}{\mathrm{d}t}\hat{\rho}_\mathrm{ex}(t)
=
-\frac{i}{\hbar}\mathcal{L}_\mathrm{ex}\hat{\rho}_\mathrm{ex}(t)
+\frac{1}{\hbar^2}\int_0^t \mathrm{d}\tau \,\mathcal{K}(t-\tau)\,\hat{\rho}_\mathrm{ex}(\tau),
\end{equation}
where
$\mathcal{L}_\mathrm{ex}\hat{\rho}_\mathrm{ex}
=
\left[\tilde{H}_\mathrm{ex}+\langle \tilde{H}_\mathrm{ex-nu}\rangle,\hat{\rho}_\mathrm{ex}\right]$,
and $\mathcal{K}(t)$ is the memory kernel. Expanding the memory kernel to  second order in  $\hat{H}^{(\mathrm{I})}$ gives
\begin{equation}
    \begin{split}
        &\mathcal{K}_{nm,kl}^{(2)}(t) = \left<\delta \hat{H}^{(\mathrm{I})}_{lm}(t) \delta \hat{H}^{(\mathrm{I})}_{nk}\right> + \left<\delta \hat{H}^{(\mathrm{I})}_{kn}(t) \delta \hat{H}^{(\mathrm{I})}_{ml}\right>^{*} \\
        &-\delta_{lm}\sum_r \left<\delta \hat{H}^{(\mathrm{I})}_{nr}(t) \delta \hat{H}^{(\mathrm{I})}_{rk}\right> - \delta_{nk}\sum_r \left<\delta \hat{H}^{(\mathrm{I})}_{mr}(t) \delta \hat{H}^{(\mathrm{I})}_{rl}\right>^{*}
    \end{split}
\end{equation}
where
$\delta\hat{H}^{(\mathrm{I})}
=
\hat{H}^{(\mathrm{I})}-\langle \hat{H}^{(\mathrm{I})}\rangle$, interaction picture $\delta\hat{H}^{(\mathrm{I})}(t)=e^{\frac{i}{\hbar}\hat{H}^{(0)}t}\delta\hat{H}^{(\mathrm{I})}e^{-\frac{i}{\hbar}\hat{H}^{(0)}t}$.

Define the correlation function
\begin{align}
    \begin{aligned}
    C_{nm,kl}(t)\equiv\left<\delta \hat{H}^{(\mathrm{I})}_{nm}(t) \delta \hat{H}^{(\mathrm{I})}_{kl}\right> = \left\langle
    \hat{H}^{(\mathrm{I})}_{nm}(t)\hat{H}^{(\mathrm{I})}_{kl}
    \right\rangle
    -
    \left\langle\hat{H}^{(\mathrm{I})}_{nm}\right\rangle
    \left\langle\hat{H}^{(\mathrm{I})}_{kl}\right\rangle
    \end{aligned}
\end{align}
The average interaction term $\left\langle\hat{H}^{(\mathrm{I})}_{nm}\right\rangle$ can be evaluated using the Bloch identity $\left\langle e^{\alpha\hat{b} + \beta\hat{b}}\right\rangle_\mathrm{nu}=e^{(1 / 2)\left\langle(\alpha\hat{b} + \beta\hat{b})^2\right\rangle_\mathrm{nu}}$, such that
\begin{equation}
    \left\langle\hat{H}^{(\mathrm{I})}_{nm}\right\rangle_\mathrm{nu} = \sum_\alpha \frac{V_{nm}^\alpha}{2\omega_\alpha^2}\left({V_{nn}^\alpha}+{V_{mm}^\alpha}\right)\exp{-\sum_\gamma{\frac{\left({V}_{nn}^\gamma-{V}_{mm}^\gamma\right)^2}{2\hbar\omega_\gamma^3}}\left(n_\gamma + \frac{1}{2}\right)}.
\end{equation}
 This term is usually exponentially small in the full polaron transformation framework.~\cite{jang2022partially} The two-point correlation function can be evaluated analytically by the finite temperature Wick theorem~\cite{izmaylov2011nonequilibrium}
\begin{align}
\begin{aligned}
        \left\langle
    \hat{H}^{(\mathrm{I})}_{nm}(t)\hat{H}^{(\mathrm{I})}_{kl}
    \right\rangle &\equiv \left[h_{nm,kl}(t)h_{lk,nm}(t) + l_{nm,kl}(t) \right]f_{nm,kl}^\mathrm{FC}(t)\\
        h_{nm,kl}(t) &= \sum_\alpha-\frac{{V}_{nm}^\alpha}{2\omega^2_\alpha}\left({V}_{kk}^\alpha-{V}_{ll}^\alpha\right)\left[n_\alpha e^{i\omega_\alpha t}-(n_\alpha+1) e^{-i\omega_\alpha t}\right]-\frac{{V}_{nm}^\alpha}{2\omega^2_\alpha}\left({V}_{nn}^\alpha+{V}_{mm}^\alpha\right)\\
    l_{nm,kl}(t) &= \frac{\hbar{V}_{nm}^\alpha{V}_{kl}^\alpha }{2\omega_\alpha}\left[n_\alpha e^{i\omega_\alpha t} + (n_\alpha+1) e^{-i\omega_\alpha t}\right] \\
    f_{nm,kl}^\mathrm{FC}(t) &= \exp \Bigg\{ \sum_\alpha\frac{1}{2\hbar\omega_\alpha^3}\left({V}_{mm}^\alpha - {V}_{nn}^\alpha\right)\left({V}_{kk}^\alpha - {V}_{ll}^\alpha\right) \left[\left({n}_\alpha + 1\right) e^{-i \omega_\alpha t} + n_\alpha e^{i \omega_\alpha t}\right] \\
& \hspace{1cm} - \frac{1}{2\hbar\omega_\alpha^3}\left[\left({V}_{mm}^\alpha - {V}_{nn}^\alpha\right)^2 + \left({V}_{kk}^\alpha - {V}_{ll}^\alpha\right)^2\right]\left(n_\alpha+\frac{1}{2}\right)\Bigg\}.
\end{aligned}
\end{align}
where the average thermal occupation $n_\alpha=1/(e^{\beta\hbar\omega_\alpha}-1)$.

\section{Frozen-Mode Quantum Master Equation}

For the frozen-mode QME (FM-QME), the low-frequency nuclear vibrations are treated explicitly as static disorder, while the remaining fast modes are treated perturbatively within the same polaron-transformed QME framework as described in the previous section. We partition each mode into slow and fast components using a frequency-dependent switching function $s_\alpha$,
\begin{equation}
    V_{nm}^{\alpha,\mathrm{s}}=\sqrt{s_\alpha}\,V_{nm}^{\alpha},
    \qquad
    V_{nm}^{\alpha,\mathrm{f}}=\sqrt{1-s_\alpha}\,V_{nm}^{\alpha},
\end{equation}
where the splitting function is
\begin{equation}
    s_\alpha =
    \begin{cases}
        \left[1-\left(\omega_\alpha/\omega_{\rm split}\right)^\nu\right]^2,
        & |\omega_\alpha| < \omega_{\rm split},
        \\
        0,
        & |\omega_\alpha| \ge \omega_{\rm split}.
    \end{cases}
\end{equation}
The parameter $\nu$ controls the sharpness of the frequency cutoff. In Fig.~2 of the main text, $\hbar\omega_{\rm split}=23$ meV and $\nu=4$.

For each trajectory $\xi$, we sample one static set of slow-mode coordinates $\bigl\{Q_\alpha^{(\xi)}\bigr\}_{\alpha=1}^{N_{\rm slow}}$
from the thermal distribution 
\begin{equation}
    Q_\alpha^{(\xi)} \sim \mathcal N\!\left(0,\frac{k_{\rm B}T}{\omega_\alpha^2}\right)
\end{equation}
and hold them fixed during the dynamics, giving a trajectory-dependent excitonic Hamiltonian
\begin{equation}
\begin{split}
    \hat{H}^{(\xi)}_\mathrm{ex}
    &=
    \hat{H}_\mathrm{ex}
    +
    \sum_{n,m=1}^{N_\mathrm{ex}}\ket{\psi_n}\bra{\psi_m}
    \sum_{\alpha=1}^{N_\mathrm{nu}}
    V_{nm}^{\alpha,\mathrm{s}} Q_\alpha^{(\xi)} \\
    &= \sum_{n,m=1}^{N_\mathrm{ex}}\ket{\psi_n}\bra{\psi_m}
    \left(E_n^0\delta_{nm}+\sum_{\alpha=1}^{N_\mathrm{nu}}
    V_{nm}^{\alpha,\mathrm{s}} Q_\alpha^{(\xi)}\right)
\end{split}
\end{equation}
We choose to work in the eigenstates of $ \hat{H}^{(\xi)}_\mathrm{ex}$ defined by
\begin{align}
     \hat{H}^{(\xi)}_\mathrm{ex}\ket{\varphi^{(\xi)}_a} = E_a^{(\xi)}\ket{\varphi^{(\xi)}_a},
\end{align}
and express the fast part of the exciton--phonon coupling in this basis as
\begin{equation}
    V_{ab}^{\alpha,\mathrm{f},(\xi)}
    =
    \sum_{n,m=1}^{N_\mathrm{ex}} \bra{\varphi_a^{(\xi)}}\ket{\psi_n}V_{nm}^{\alpha,\mathrm{f}}\bra{\psi_m}
    \ket{\varphi_b^{(\xi)}}.
\end{equation}
The perturbative QME is then solved for each realization using
$\hat{H}^{(\xi)}_\mathrm{ex}$ as the system Hamiltonian and only the fast couplings $V_{nm}^{\alpha,\mathrm{f},(\xi)}$ in the memory kernel,
\begin{equation}
    \frac{\mathrm{d}}{\mathrm{d}t}\hat{\rho}^{(\xi)}_\mathrm{ex}(t)
    =
    -\frac{i}{\hbar}\mathcal{L}^{(\xi)}_\mathrm{ex}\hat{\rho}^{(\xi)}_\mathrm{ex}(t)
    +
    \frac{1}{\hbar^2}
    \int_0^t d\tau\,
    \mathcal{K}^{(2),(\xi)}_\mathrm{f}(t-\tau)
    \hat{\rho}^{(\xi)}_\mathrm{ex}(\tau).
\end{equation}
The final FM-QME result is obtained by averaging over the frozen-disorder ensemble,
\begin{equation}
    \hat{\rho}^{\mathrm{FM-QME}}_\mathrm{ex}(t)
    =
    \frac{1}{N_\mathrm{traj}}
    \sum_{\xi=1}^{N_\mathrm{traj}}
    \hat{U}^{(\xi)}
    \hat{\rho}^{(\xi)}_\mathrm{ex}(t)
    \hat{U}^{(\xi)\dagger},
\end{equation}
where
$\hat{U}^{(\xi)}=\sum_a\lvert \varphi_a^{(\xi)} \rangle\langle \psi_a \rvert$
denotes the transformation from the trajectory-dependent eigenbasis to the diabatic basis and $N_\mathrm{traj}=10^3$. Thus, the slow modes contribute inhomogeneous static fluctuations of the excitonic Hamiltonian, while the fast modes provide dynamical relaxation through the perturbative memory kernel.

\section{Mixed Quantum--Classical Dynamics}
\label{sec:si_mqc}

All mixed quantum--classical (MQC) calculations use the classical limit of the nuclear degrees of freedom in the same linear vibronic Hamiltonian as in Eq.~\eqref{eq:si_lvc_hamiltonian}. It is useful to separate the nuclear potential from the excitonic matrix,
\begin{align}
    \hat{U}(\mathbf Q)
    &=
    U_{\rm nu}(\mathbf Q)
    +
    \hat{U}_{\rm ex}(\mathbf Q),
    \\
    U_{\rm nu}(\mathbf Q)
    &=
    \sum_{\alpha=1}^{N_{\rm nu}}
    \frac{1}{2}\omega_\alpha^2Q_\alpha^2,
    \\
    \hat{U}_{\rm ex}(\mathbf Q)
    &=
    \sum_{n,m=1}^{N_{\rm ex}}
    U^\mathrm{ex}_{nm}(\mathbf Q)\ket{\psi_n}\bra{\psi_m},
    \qquad
    U^\mathrm{ex}_{nm}(\mathbf Q)
    =
    E_n^0\delta_{nm}
    +
    \sum_{\alpha=1}^{N_{\rm nu}}V_{nm}^{\alpha}Q_\alpha .
\end{align}
The adiabatic excitonic states are defined by
\begin{equation}
    \hat {U}(\mathbf Q)\ket{\phi_a(\mathbf Q)}
    =
    U_a(\mathbf Q)\ket{\phi_a(\mathbf Q)}.
\end{equation}
The electronic wavefunction along a trajectory may be expanded either in the diabatic or adiabatic basis,
\begin{equation}
    \ket{\psi_{\rm ex}(t)}
    =
    \sum_{n=1}^{N_{\rm ex}}c_n(t)\ket{\psi_n}
    =
    \sum_{a=1}^{N_{\rm ex}}c_a(t)\ket{\phi_a(\mathbf Q(t))}.
\end{equation}
In the diabatic basis the coefficients obey
\begin{equation}
    \dot c_n(t)
    =
    -\frac{i}{\hbar}\sum_{m=1}^{N_{\rm ex}}U_{nm}(\mathbf Q(t))c_m(t),
\end{equation}
whereas in the adiabatic basis
\begin{equation}
    \dot c_a(t)
    =
    -\frac{i}{\hbar}U_a(\mathbf Q(t))c_a(t)
    -
    \sum_{b=1}^{N_{\rm ex}}
    \sum_{\alpha=1}^{N_{\rm nu}}
    P_\alpha(t)d_{ab}^{\alpha}(\mathbf Q(t))c_b(t),
    \label{eq:si_adia_coeff}
\end{equation}
with derivative couplings
\begin{equation}
    d_{ab}^{\alpha}(\mathbf Q)
    =
    \bra{\phi_a(\mathbf Q)}
    \frac{\partial}{\partial Q_\alpha}
    \ket{\phi_b(\mathbf Q)}
    =
    \frac{
    \bra{\phi_a(\mathbf Q)}
    \partial \hat{U}_{\rm ex}(\mathbf Q)/\partial Q_\alpha
    \ket{\phi_b(\mathbf Q)}
    }
    {U_b(\mathbf Q)-U_a(\mathbf Q)}
    \qquad (a\ne b).
\end{equation}
The nuclear equations have the common form
\begin{equation}
    \dot Q_\alpha=P_\alpha,
    \qquad
    \dot P_\alpha=F_\alpha(\mathbf Q),
\end{equation}
with the force \(F_\alpha\) specified by the particular MQC approximation. The initial conditions for the nuclear degrees of freedom are sampled from the classical Boltzmann distribution of the uncoupled harmonic bath,
\begin{equation}
\rho_{\rm nu}^{\rm cl}(\mathbf Q,\mathbf P)
\propto
\exp\left[
-\beta
\sum_{\alpha=1}^{N_{\rm nu}}
\left(
\frac{P_\alpha^2}{2}
+
\frac{1}{2}\omega_\alpha^2Q_\alpha^2
\right)
\right].
\end{equation}
Depending on the MQC methods, the observables, such as populations and exciton energy $U_\mathrm{ex}(t)$, are well converged with $10^4$ trajectories. 

\subsection{Mean-field Ehrenfest dynamics}
In mean-field Ehrenfest dynamics the nuclei evolve on the electronic expectation value of the force operator.~\cite{Tully1998,Kapral2006}
\begin{align}
    F_\alpha^{\rm MF}(\mathbf Q)
    &=
    -
    \bra{\psi_{\rm ex}(t)}
    \frac{\partial \hat U(\mathbf Q)}{\partial Q_\alpha}
    \ket{\psi_{\rm ex}(t)}
    \nonumber\\
    &=
    -\omega_\alpha^2Q_\alpha
    -
    \sum_{n,m=1}^{N_{\rm ex}}
    c_n^*(t)c_m(t)V_{nm}^{\alpha},
    \label{eq:si_ehrenfest_force}
\end{align}
where the force is a coherent average over all occupied excitonic states.

\subsection{Spin-LSC dynamics}

The spin linearized semiclassical (spin-LSC) method maps the \(N_{\rm ex}\)-state excitonic subsystem to continuous spin-mapping variables while preserving the SU(\(N_{\rm ex}\)) structure of the electronic Hilbert space.~\cite{Runeson2019,Runeson2020} We use the W-representation, for which the real mapping variables \(\{q_n,p_n\}_{n=1}^{N_{\rm ex}}\) are constrained to the hypersphere
    \begin{equation}
    \sum_{n=1}^{N_{\rm ex}}\left(q_n^2+p_n^2\right)
    =
    R_{\rm W}^2,
    \qquad
    R_{\rm W}^2=2\sqrt{N_{\rm ex}+1},
\end{equation}
and the zero-point-energy parameter is
\begin{equation}
    \gamma_{\rm W}
    =
    \frac{2}{N_{\rm ex}}
    \left(\sqrt{N_{\rm ex}+1}-1\right).
\end{equation}
For the real symmetric excitonic Hamiltonian used here, the spin-mapping Hamiltonian is
\begin{align}
    H_{\rm W}
    &=
    \sum_{\alpha=1}^{N_{\rm nu}}
    \left[
    \frac{P_\alpha^2}{2}
    +
    \frac{1}{2}\omega_\alpha^2Q_\alpha^2
    \right]
    +
    \frac{1}{2}\sum_{n,m}^{N_{\rm ex}}
    U_{nm}^\mathrm{ex}(\mathbf Q)
    \left(q_nq_m+p_np_m-\gamma_{\rm W}\delta_{nm}\right).
    \label{eq:si_spin_lsc_hamiltonian}
\end{align}
The electronic mapping variables are propagated according to
\begin{equation}
    \dot q_n
    =
    \frac{1}{\hbar}\sum_{m=1}^{N_{\rm ex}}U_{nm}^\mathrm{ex}(\mathbf Q)p_m,
    \qquad
    \dot p_n
    =
    -\frac{1}{\hbar}\sum_{m=1}^{N_{\rm ex}}U_{nm}^\mathrm{ex}(\mathbf Q)q_m,
    \label{eq:si_spin_lsc_eom}
\end{equation}
while the nuclear force is
\begin{align}
    F_\alpha^{\rm spin-LSC}(\mathbf{Q})
    &=
    -\frac{\partial H_{\rm W}}{\partial Q_\alpha}
    \nonumber\\
    &=
    -\omega_\alpha^2Q_\alpha-
    \frac{1}{2}\sum_{n,m}^{N_{\rm ex}}
    V_{nm}^\alpha
    \left(q_nq_m+p_np_m-\gamma_{\rm W}\delta_{nm}\right).
    \label{eq:si_spin_lsc_force}
\end{align}
The W-representation estimator for the diabatic population of state \(n\) is
\begin{equation}
    \Pi_n^{\rm W}
    =
    \frac{1}{2}\left(q_n^2+p_n^2-\gamma_{\rm W}\right),
    \qquad
    \mathcal P_n^{(\rm d)}(t)
    =
    \left\langle \Pi_n^{\rm W}(t)\right\rangle_{\rm traj}.
\end{equation}
For an initial excitation in diabatic state \(e\), focused spin-mapping initial conditions are generated from
\begin{equation}
    q_n=r_n\cos\varphi_n,
    \qquad
    p_n=r_n\sin\varphi_n,
    \qquad
    r_e=\sqrt{2+\gamma_{\rm W}},
    \qquad
    r_{n\ne e}=\sqrt{\gamma_{\rm W}},
\end{equation}
with phases \(\varphi_n\) sampled uniformly on \([0,2\pi)\).

\subsection{Fewest-switches surface hopping}

Fewest-switches surface hopping (FSSH) propagates the nuclei on a single active adiabatic surface labelled by \(a(t)\), while the electronic amplitudes evolve according to Eq.~\eqref{eq:si_adia_coeff}.~\cite{Tully1990,Subotnik2016} The force on the active surface is 
\begin{align}
    F_\alpha^{\mathrm{SH}}(\mathbf{Q})
    &=
    -\frac{\partial U_a(\mathbf Q)}{\partial Q_\alpha}
    =
    -\omega_\alpha^2Q_\alpha
    -
    \bra{\phi_a(\mathbf Q)}
    \frac{\partial \hat U_{\rm ex}(\mathbf Q)}{\partial Q_\alpha}
    \ket{\phi_a(\mathbf Q)}.
\end{align}
During a time step \(\Delta t\), the probability of a hop from the active state \(a\) to another adiabatic state \(b\) is
\begin{equation}
    g_{a\to b}
    =
    \max\left[
    0,\,
    \frac{2\Delta t}{|c_a|^2}
    \operatorname{Re}
    \left\{
    c_a^*c_b
    \sum_{\alpha=1}^{N_{\rm nu}}
    P_\alpha d_{ab}^{\alpha}(\mathbf Q)
    \right\}
    \right].
    \label{eq:si_fssh_prob}
\end{equation}
If a hop is accepted, the nuclear momentum is adjusted along the derivative-coupling direction to conserve the total classical energy,
\begin{equation}
    \frac{1}{2}\left|\mathbf P+\eta\mathbf d_{ab}\right|^2
    +
    U_b(\mathbf Q)
    =
    \frac{1}{2}\left|\mathbf P\right|^2
    +
    U_a(\mathbf Q),
    \label{eq:si_fssh_rescale}
\end{equation}
where \(\mathbf d_{ab}\) has components \(d_{ab}^{\alpha}\). If Eq.~\eqref{eq:si_fssh_rescale} has no real solution, the hop is rejected as frustrated.
The corresponding active-surface estimator gives adiabatic populations as \(\Pi_b^{\rm FSSH}=\delta_{b,a(t)}\).
In the diabatic basis, we use the
standard mixed estimator rather than projecting only the active adiabatic
state. Defining
\(\mathcal U_{na}(\mathbf Q)=\braket{\psi_n}{\phi_a(\mathbf Q)}\), the
diabatic population of state \(n\) is evaluated as
\begin{equation}
    \Pi_n^{\rm FSSH}(t)
    =
    \left|\mathcal U_{n a}(\mathbf Q)\right|^2
    +
    \operatorname{Re}
    \left[
    \sum_a\sum_{b\neq a}
    c_a^*(t)\mathcal U_{na}^*(\mathbf Q)
    \mathcal U_{nb}(\mathbf Q)c_b(t)
    \right].
\end{equation}
The first term is the active-surface contribution, while the second term
retains the off-diagonal adiabatic coherences. No decoherence scheme is applied. 

\subsection{Mapping approach to surface hopping}

The mapping approach to surface hopping (MASH) uses the same adiabatic electronic equation of motion and the same single-surface force as surface hopping, but the active surface is selected deterministically from the electronic amplitudes rather than stochastically.~\cite{Mannouch2023,Runeson2023} At each time,
\begin{equation}
    a(t)
    =
    \operatorname*{argmax}_{b}
    |c_b(t)|^2,
    \label{eq:si_mash_active}
\end{equation}
where the coefficients are expressed in the instantaneous adiabatic basis. A hop occurs when the identity of the largest adiabatic population changes. For a hop \(a\to b\), the MASH momentum rescaling direction is determined by the gradient of the adiabatic population difference,
and the momentum is rescaled to conserve energy. If there is insufficient kinetic energy, this component is reversed and the hop is rejected.

For observables, the multi-state MASH estimator is not simply the active-state indicator except for adiabatic equilibrium populations. For a chosen electronic basis \(\{\ket{\psi_n}\}_{n=1}^{N_{\rm ex}}\), with amplitudes \(c_n=\braket{\psi_n}{\psi_{\rm ex}}\), the population estimator is
\begin{equation}
    \Phi_n
    =
    \frac{1}{N_{\rm ex}}
    +
    \alpha_{N_{\rm ex}}
    \left(
    |c_n|^2-\frac{1}{N_{\rm ex}}
    \right),
    \qquad
    \alpha_{N_{\rm ex}}
    =
    \frac{N_{\rm ex}-1}{H_{N_{\rm ex}}-1},
    \qquad
    H_{N_{\rm ex}}
    =
    \sum_{m=1}^{N_{\rm ex}}\frac{1}{m}.
    \label{eq:si_mash_population_estimator}
\end{equation}
More generally, for an electronic operator \(\hat O=\sum_{n,m}O_{nm}\ket{\psi_n}\bra{\psi_m}\), the trajectory estimator is
\begin{equation}
    O_{\rm MASH}
    =
    \sum_{n,m=1}^{N_{\rm ex}}
    O_{nm}
    \left[
    \alpha_{N_{\rm ex}}c_n^*c_m
    +
    \beta_{N_{\rm ex}}\delta_{nm}
    \right],
    \qquad
    \beta_{N_{\rm ex}}
    =
    \frac{1-\alpha_{N_{\rm ex}}}{N_{\rm ex}}.
    \label{eq:si_mash_observable_estimator}
\end{equation}
For a pure initial state \(\ket{\psi_i}\), the electronic amplitudes are sampled uniformly on the normalized complex sphere subject to the condition that \(|c_i|^2\) is the largest population, which gives \(\langle\Phi_n(0)\rangle=\delta_{ni}\). This sampling, together with the deterministic active-surface rule in Eq.~\eqref{eq:si_mash_active}, gives the correct quantum--classical equilibrium populations in the long-time ergodic limit.

\section{Diabatic Thermal Population}
The diabatic thermal equilibrium population of state \(n\) is defined as
\begin{equation}
    \mathcal{P}^{(\rm d)}_{n, {\rm eq}} = \frac{\mathrm{Tr}\left\{e^{-\beta\hat{H}}\ket{\psi_n}\bra{\psi_n}\right\}}{\mathrm{Tr}\left\{e^{-\beta\hat{H}}\right\}}.
\end{equation}
First, rewrite the total Hamiltonian as 
\begin{equation}
    \hat{H}\equiv \hat{T} + {U}(\hat{\mathbf{Q}})
\end{equation}
where
\begin{align}
    \hat{T} &= \sum_{\alpha}\frac{1}{2}\hat{P}_{\alpha}^2 \\
    {U}(\hat{\mathbf{Q}}) &= \sum_{\alpha=1}^{N_\mathrm{nu}}\frac{1}{2}\omega_\alpha^2 \hat{Q}_\alpha^2+\sum_{n,m=1}^{N_\mathrm{ex}}\ket{\psi_n}\bra{\psi_m}\left(\sum_{\alpha=1}^{N_\mathrm{nu}} V_{nm}^\alpha\hat{Q}_{\alpha} + \delta_{nm}E_n^0\right).
\end{align}
Performing Trotter factorization with $N_p$ points
\begin{align}
    e^{-\beta(\hat{T}+\hat{U})} = \lim_{N_p\to\infty}\left(e^{-\epsilon\hat{T}}e^{-\epsilon\hat{U}}\right)^{N_p}
\end{align}
where $\epsilon=\beta/N_p$. Inserting the identity operators $\int\mathrm{d}\mathbf{Q}_i\ket{\mathbf{Q}_i}\bra{\mathbf{Q}_i}$ between each Trotter term such that the partition function can be written as 
\begin{align*}
    \operatorname{Tr}\left[e^{-\beta \hat{H}}\right]
    &=\lim _{N_p \rightarrow \infty} \left(\frac{1}{2 \pi \hbar^2 \epsilon}\right)^{N_p N_{\mathrm{nu}} / 2} \int \mathrm{d}^{N_p}\mathbf{Q}\exp\left[-\frac{N_p}{2 \hbar^2 \beta}\sum_{i=1}^{N_p}\sum_{\alpha=1}^{N_\mathrm{nu}}({Q}_{\alpha,i}-{Q}_{\alpha,i+1})^2\right]\\
    &\times\operatorname{Tr}_{\mathrm{ex}}\left\{\prod_{i=1}^{N_p} e^{-\frac{\beta}{N_p} \hat{U}\left(\mathbf{Q}_i\right)}\right\}
\end{align*}
where $\mathrm{d}^{N_p}\mathbf{Q} = \prod_\alpha^{N_{\mathrm{nu}}}\prod_i^{N_p}\mathrm{d}Q_{\alpha,i}$ and $Q_{\alpha,N_p+1}=Q_{\alpha,1}$. Similarly, one would have
\begin{align}
    \begin{aligned}
    &\operatorname{Tr}\left[e^{-\beta \hat{H}}\ket{\psi_n}\bra{\psi_n}\right]\\
    &=\lim _{N_p \rightarrow \infty} \left(\frac{1}{2 \pi \hbar^2 \epsilon}\right)^{N_p N_{\mathrm{nu}} / 2} \int \mathrm{d}^{N_p}\mathbf{Q} \exp\left[-\frac{N_p}{2 \hbar^2 \beta}\sum_{i=1}^{N_p}\sum_{\alpha=1}^{N_\mathrm{nu}}({Q}_{\alpha,i}-{Q}_{\alpha,i+1})^2\right]\\
    &\times\operatorname{Tr}_{\mathrm{ex}}\left\{\prod_{i=1}^{N_p} e^{-\frac{\beta}{N_p} \hat{U}\left(\mathbf{Q}_i\right)}\ket{\psi_n}\bra{\psi_n}\right\}.
    \end{aligned}
\end{align}

Following the convention of ring-polymer formalism, we partition the harmonic potential with the nuclear kinetic energy and define
\begin{align}
    U_\mathrm{RP}(\left\{Q_{\alpha,i}\right\}) = \sum_{\alpha=1}^{N_{\mathrm{nu}}}\sum_{i=1}^{N_p}\left[\frac{1}{2}\omega_p^2\left({Q}_{\alpha,i}-{Q}_{\alpha, i+1}\right)^2 + \frac{1}{2}\omega_\alpha^2Q_{\alpha,i}^2\right]
\end{align}
where $\omega_p = 1/(\epsilon\hbar) = N_p/(\beta\hbar)$ is the spring frequency between the coupled beads coordinates $\left\{Q_{\alpha,i}\right\}_{i=1}^{N_p}$. The remaining excitonic potential can be written as 
\begin{align}
    \hat{U}_{\rm ex}(\mathbf{Q}_i) = \sum_{n,m}^{N_{\rm ex}}\ket{\psi_n}\bra{\psi_m}\left( \delta_{nm}E_n^0+\sum_{\alpha}^{N_{\mathrm{nu}}} V_{nm}^\alpha{Q}_{\alpha,i}\right).
\end{align}
The equilibrium population can then be written as
\begin{align}\label{eq:pimc_population_estimator}
    \mathcal{P}^{(\rm d)}_{n, {\rm eq}} &= \lim _{N_p \rightarrow \infty}\dfrac{\int \mathrm{d}^{N_p}\mathbf{Q} \, e^{-\epsilon U_\mathrm{RP}(\left\{Q_{\alpha,i}\right\})} \bra{\psi_n}\prod_{i=1}^{N_p}e^{-\epsilon \hat{U}_{\rm ex}(\mathbf{Q}_i)}\ket{\psi_n}}{\int \mathrm{d}^{N_p}\mathbf{Q} \, e^{-\epsilon U_\mathrm{RP}(\left\{Q_{\alpha,i}\right\})}  \operatorname{Tr}_{\mathrm{ex}}\left\{\prod_{i=1}^{N_p}e^{-\epsilon \hat{U}_{\rm ex}(\mathbf{Q}_i)}\right\}}
\end{align}

Equation~\eqref{eq:pimc_population_estimator} has the form of an importance-sampling average. In practice, we sample nuclear ring-polymer configurations \(\{Q_{\alpha,i}^{(s)}\}\) from the Gaussian distribution proportional to
\[
    \exp\left[-\epsilon U_\mathrm{RP}(\{Q_{\alpha,i}\})\right].
\]
This sampling can be performed efficiently by transforming the bead coordinates to ring-polymer normal modes, where \(U_\mathrm{RP}\) is diagonal and each normal-mode coordinate is sampled independently from a Gaussian distribution. After transforming back to bead coordinates, we construct \(\hat{U}_{\rm ex}(\mathbf{Q}_i^{(s)})\) at each bead and evaluate
\begin{align}
    \hat{R}_s
    &=
    \prod_{i=1}^{N_p}
    e^{-\epsilon \hat{U}_{\rm ex}(\mathbf{Q}_i^{(s)})},
    \\
    W_s
    &=
    \operatorname{Tr}_{\rm ex}\{\hat{R}_s\},
    \\
    p_{n,s}
    &=
    \frac{
    \bra{\psi_n}\hat{R}_s\ket{\psi_n}
    }{
    W_s
    }.
\end{align}
The diabatic equilibrium population is then estimated from \(N_{\rm s}\) Monte Carlo samples using $N_\mathrm{s}=10^6$ samples
\begin{equation}
    \mathcal{P}^{(\rm d)}_{n,{\rm eq}}
    \approx
    \frac{
    \sum_{s=1}^{N_{\rm s}} W_s p_{n,s}
    }{
    \sum_{s=1}^{N_{\rm s}} W_s
    }
    =
    \frac{
    \sum_{s=1}^{N_{\rm s}}
    \bra{\psi_n}\hat{R}_s\ket{\psi_n}
    }{
    \sum_{s=1}^{N_{\rm s}}
    \operatorname{Tr}_{\rm ex}\{\hat{R}_s\}
    }.
    \label{eq:pimc_mc_estimator}
\end{equation}

\section{Classical Fermi's Golden Rule in Adiabatic Basis}

For a two-level linear vibronic Hamiltonian, we write the diabatic
representation in the basis $\{\ket{\psi_1},\ket{\psi_2}\}$ as
\begin{equation}
\begin{split}
\hat{\mathbf H}_{\rm dia}
&=
\left[
\sum_{\alpha=1}^{N_{\rm nu}}
\frac{1}{2}
\left(
\hat P_\alpha^2+\omega_\alpha^2\hat Q_\alpha^2
\right)
\right]\mathbf I_2
+
\begin{pmatrix}
E_1+\sum_{\alpha=1}^{N_{\rm nu}}V_{11}^\alpha \hat Q_\alpha
&
\sum_{\alpha=1}^{N_{\rm nu}}V_{12}^\alpha \hat Q_\alpha
\\
\sum_{\alpha=1}^{N_{\rm nu}}V_{12}^\alpha \hat Q_\alpha
&
E_2+\sum_{\alpha=1}^{N_{\rm nu}}V_{22}^\alpha \hat Q_\alpha
\end{pmatrix}
\\
&=
\frac{\hat{\mathbf P}^2}{2}\mathbf I_2
+
\bar V(\hat{\mathbf Q})\mathbf I_2
+
\begin{pmatrix}
-D(\hat{\mathbf Q}) & W(\hat{\mathbf Q})\\
W(\hat{\mathbf Q}) & D(\hat{\mathbf Q})
\end{pmatrix}\equiv
\frac{\hat{\mathbf P}^2}{2}\mathbf I_2
+
\mathbf U_{\rm ex}(\hat{\mathbf Q}),
\end{split}
\label{eq:two_level_dia_ham}
\end{equation}
where the nuclear coordinates are assumed to be mass-weighted, and
\begin{align}
\bar V(\mathbf Q)
&=
\frac{E_1+E_2}{2}
+
\sum_{\alpha=1}^{N_{\rm nu}}
\left[
\frac{1}{2}\omega_\alpha^2 Q_\alpha^2
+
\frac{V_{11}^\alpha+V_{22}^\alpha}{2}Q_\alpha
\right],
\label{eq:barV_def}
\\
D(\mathbf Q)
&=
\frac{E_2-E_1}{2}
+
\sum_{\alpha=1}^{N_{\rm nu}}
\frac{V_{22}^\alpha-V_{11}^\alpha}{2}Q_\alpha,
\label{eq:D_def}
\\
W(\mathbf Q)
&=
\sum_{\alpha=1}^{N_{\rm nu}}V_{12}^\alpha Q_\alpha .
\label{eq:W_def}
\end{align}

The adiabatic excitonic states are defined by diagonalizing the
coordinate-dependent potential matrix at fixed nuclear configuration,
\begin{equation}
\mathbf U_{\rm ex}(\mathbf Q)\ket{\phi_a(\mathbf Q)}
=
U_a^{\rm ex}(\mathbf Q)\ket{\phi_a(\mathbf Q)} .
\label{eq:ad_eigenproblem}
\end{equation}
The corresponding adiabatic potential energy surfaces are
\begin{align}
U_1^{\rm ex}(\mathbf Q)
&=
\bar V(\mathbf Q)
-
\sqrt{D^2(\mathbf Q)+W^2(\mathbf Q)},
\\
U_2^{\rm ex}(\mathbf Q)
&=
\bar V(\mathbf Q)
+
\sqrt{D^2(\mathbf Q)+W^2(\mathbf Q)} .
\end{align}

In the adiabatic basis $\left\{\ket{\phi_1(\mathbf{Q})}, \ket{\phi_2(\mathbf{Q})}\right\}$, the nuclear kinetic energy generates
derivative-coupling terms. The Hamiltonian becomes~\cite{jang2025fermi}
\begin{equation}
\hat{\mathbf H}_{\rm ad}
=
\frac{\hat{\mathbf P}^2}{2}\mathbf I_2
+
\begin{pmatrix}
U_1^{\rm ex}(\hat{\mathbf Q})+\hbar^2\hat\Lambda_{11}
&
\hbar^2\hat\Lambda_{12}
\\
\hbar^2\hat\Lambda_{21}
&
U_2^{\rm ex}(\hat{\mathbf Q})+\hbar^2\hat\Lambda_{22}
\end{pmatrix},
\label{eq:ad_ham}
\end{equation}
where $\hat\Lambda_{ab}$ is the derivative coupling operator defined by
\begin{equation}
\hat\Lambda_{ab}
=
-\mathbf d_{ab}(\mathbf Q)\cdot\nabla_{\mathbf Q}
-\frac{1}{2}g_{ab}(\mathbf Q).
\label{eq:Lambda_def}
\end{equation}
Here, the diagonal and off-diagonal first-order derivative couplings are
\begin{align}
    d_{ab}^\alpha(\mathbf{Q}) &\equiv \bra{\phi_a(\mathbf{Q})} \nabla_{Q_\alpha}\ket{\phi_b(\mathbf{Q})} \\
    d_{11}^\alpha(\mathbf{Q}) &= d_{22}^\alpha(\mathbf{Q}) = 0\\
    d_{12}^\alpha(\mathbf{Q}) &= -d_{21}^\alpha(\mathbf{Q}) = \frac{V_{12}^\alpha D(\mathbf{Q})-\frac{1}{2}{(V_{22}^\alpha-V_{11}^\alpha)}W(\mathbf{Q})}{2\left[W^2(\mathbf{Q}) + D^2(\mathbf{Q})\right]}.
\end{align}
and the second-order derivatives are
\begin{equation}
g_{ab}(\mathbf Q)\equiv \bra{\phi_a(\mathbf{Q})}\nabla^2_\mathbf{Q}\ket{\phi_b(\mathbf{Q})}
=
\nabla_{\mathbf Q}\cdot \mathbf d_{ab}(\mathbf Q)
+
\sum_k
\mathbf d_{ak}(\mathbf Q)\cdot \mathbf d_{kb}(\mathbf Q).
\label{eq:g_identity}
\end{equation}
For the two-level model, this gives
\begin{align}
\hat\Lambda_{11}
&=
\hat\Lambda_{22}
=
\frac{1}{2}
\norm{\mathbf d_{12}(\mathbf Q)}^2,
\\
\hat\Lambda_{ab}
&=
-\mathbf d_{ab}(\mathbf Q)\cdot\nabla_{\mathbf Q}
-\frac{1}{2}
\nabla_{\mathbf Q}\cdot\mathbf d_{ab}(\mathbf Q), \quad a\neq b.
\end{align}

Finally, the off-diagonal derivative-coupling operator can be written in a symmetrized momentum-coupling form. Define
\begin{equation}
\mathbf A_{ab}(\mathbf Q)
\equiv
-i\hbar\,\mathbf d_{ab}(\mathbf Q),
\label{eq:A_nm_def}
\end{equation}
The following equality is satisfied for an arbitrary nuclear
wavefunction $\chi(\mathbf Q)$,
\begin{equation}
\begin{split}
\frac{1}{2}
\left(
\hat{\mathbf{A}}_{ab}\cdot\hat{\mathbf{P}} + \hat{\mathbf{P}}\cdot \hat{\mathbf{A}}_{ab}
\right)\chi(\mathbf Q)
&=
-i\hbar
\left[
\mathbf A_{ab}(\mathbf Q)\cdot\nabla_{\mathbf Q}\chi(\mathbf Q)
+
\frac{1}{2}
\left(
\nabla_{\mathbf Q}\cdot\mathbf A_{ab}(\mathbf Q)
\right)
\chi(\mathbf Q)
\right]
\\
&=
\hbar^2\hat\Lambda_{ab}\chi(\mathbf Q).
\end{split}
\end{equation}
Therefore,
\begin{equation}
\hbar^2\hat\Lambda_{ab}
=
\frac{1}{2}
\left(
\hat{\mathbf{A}}_{ab}\cdot\hat{\mathbf{P}} + \hat{\mathbf{P}}\cdot \hat{\mathbf{A}}_{ab}
\right).
\label{eq:Lambda12_symmetrized}
\end{equation}

To calculate the transition rate between those two adiabatic surfaces, define the zeroth-order and interaction Hamiltonian to be
\begin{equation}
    \begin{split}
        \hat{H}^{(0)} &= \frac{\hat{\mathbf{P}}^2}{2} + \sum_a U_a^{\rm ex}(\hat{\mathbf{Q}})\ket{\phi_a({\mathbf{Q}})}\bra{\phi_a({\mathbf{Q}})}\equiv \sum_a\hat{H}_a\ket{\phi_a({\mathbf{Q}})}\bra{\phi_a({\mathbf{Q}})}\\
        \hat{H}^{(\mathrm{I})} &\equiv \sum_{a\neq b} \hbar^2\hat{\Lambda}_{ab}\ket{\phi_a({\mathbf{Q}})}\bra{\phi_b({\mathbf{Q}})}
    \end{split}
\end{equation}
where $\hat{H}_a \equiv  \dfrac{\hat{\mathbf{P}}^2}{2} + U_a^{\rm ex}(\hat{\mathbf{Q}})$. A general Fermi's Golden-Rule expression for a two-state problem is given by
\begin{equation}
\begin{split}
    k_{a \leftarrow b}&=\frac{1}{\hbar^2} \int_{-\infty}^{\infty} \mathrm{d} t \operatorname{Tr}_{\rm nu}\left\{\hat{\rho}(0) e^{\frac{i}{\hbar} \hat{H}^{(0)} t} \hat{H}^{(\mathrm{I})} e^{-\frac{i}{\hbar} \hat{H}^{(0)} t} \hat{H}^{(\mathrm{I})}\right\}\\
    &=\frac{1}{\hbar^2} \int_{-\infty}^{\infty} \mathrm{d} t \operatorname{Tr}_{\rm nu}\left\{\hat\rho_{b} e^{\frac{i}{\hbar} \hat{H}_b t} \hbar^2\hat{\Lambda}_{ba} e^{-\frac{i}{\hbar} \hat{H}_a t} \hbar^2\hat{\Lambda}_{ab}\right\}
\end{split}
\end{equation}
where the initial condition $\hat{\rho} = \ket{\phi_b}\bra{\phi_b}\otimes \hat\rho_{b}$ with the nuclear equilibrated at the initial surface. The energy gap Hamiltonian is the form of
\begin{align}
    \hat{H}_a = \hat{H}_b + \Delta \hat U, \hspace{0.5 cm} \Delta \hat U = U_a^{\rm ex}(\hat{\mathbf{Q}}) - U_b^{\rm ex}(\hat{\mathbf{Q}}).
\end{align}
One can then work in the interaction picture of $\hat{H}_b$ and define
\begin{equation}
\hat \Lambda_{ba}(t) \equiv e^{\frac{i}{\hbar} \hat H_b t} \hat \Lambda_{ba} e^{-\frac{i}{\hbar} \hat H_b t}, \quad \Delta \hat U(t) \equiv e^{\frac{i}{\hbar} \hat H_b t} \Delta \hat{U} e^{-\frac{i}{\hbar} \hat H_b t},
\end{equation}
such that
\begin{equation}
e^{\frac{i}{\hbar} \hat H_b t} e^{-\frac{i}{\hbar} \hat H_a t}=\mathcal{T} \exp \left[-\frac{i}{\hbar} \int_0^t \mathrm{d} \tau \Delta \hat U(\tau)\right].
\end{equation}
One could then write the FGR as
\begin{equation}
\begin{split}
    k_{a \leftarrow b}&=\frac{1}{\hbar^2} \int_{-\infty}^{\infty} \mathrm{d} t\left\langle \hbar^2\hat{\Lambda}_{ba}(t) \mathcal{T} \exp \left[-\frac{i}{\hbar} \int_0^t \mathrm{d} \tau \Delta \hat U(\tau)\right] \hbar^2\hat{\Lambda}_{ab}(0)\right\rangle_{\rm{eq}, b}
\end{split}
\end{equation}

In the dynamical classical limit, the quantum correlation function is
approximated by an average over classical trajectories propagated on the
initial adiabatic surface $U_b^{\rm ex}(\mathbf Q)$. The symmetrized
momentum-coupling operator is replaced by its classical counterpart,~\cite{neria1991quantum}
\begin{equation}
    \hbar^2\hat \Lambda_{ab}=
\frac{1}{2}
\left(
\hat{\mathbf{A}}_{ab}\cdot\hat{\mathbf{P}} + \hat{\mathbf{P}}\cdot \hat{\mathbf{A}}_{ab}
\right)\approx {\mathbf{A}}_{ab}\cdot{\mathbf{P}}.
\end{equation}
The classical golden-rule rate then becomes~\cite{egorov_vibronic_1998}
\begin{equation}
k_{a \leftarrow b}^{(\mathrm{cl})}=\frac{1}{\hbar^2} \int_{-\infty}^{\infty} \mathrm{d} t\left\langle\left[\mathbf{A}_{ba}(\mathbf{Q}(t)) \cdot \mathbf{P}(t)\right] \exp \left[-\frac{i}{\hbar} \int_0^t \mathrm{d} \tau \Delta U(\mathbf{Q}(\tau))\right]\left[\mathbf{A}_{ab}(\mathbf{Q}(0)) \cdot \mathbf{P}(0)\right]\right\rangle_{\mathrm{cl},b}.
\end{equation}

To summarize the procedure for calculating the transition rate in the adiabatic limit, we have:
\begin{enumerate}
    \item Sample the initial momentum and coordinate of each mode $(\mathbf{P}(0), \mathbf{Q}(0))$ based on the classical distribution of $\rho_b$.
    \item Propagate the trajectory according to Newton's equation on the initial surface $U_b^{\rm ex}(\mathbf{Q})$    \begin{equation}
    \dot{\mathbf{Q}}(t)=\mathbf{P}(t), \quad \dot{\mathbf{P}}(t)=-\nabla_{\mathbf{Q}} U_b^{\mathrm{ex}}(\mathbf{Q}(t)) .
    \end{equation}
    Evaluate $\mathbf{A}_{ba}(t)\cdot\mathbf{P}(t)$, $\Delta U(\mathbf{Q})$ etc.
    \item Average over the trajectories and evaluate $k_{a \leftarrow b}^{(\mathrm{cl})}$ in the dynamical classical limit. 
\end{enumerate}

The population dynamics shown in Fig.~4 of the main text is obtained from
\begin{equation}
\frac{\mathrm{d} p_b(t)}{\mathrm{d}t}
=
\sum_{\substack{a\\a\neq b}}
p_ak_{b\leftarrow a}(t)
-
\sum_{\substack{a\\a\neq b}}
p_b(t)k_{a\leftarrow b}.
\end{equation}
To ensure relaxation to the adiabatic thermal equilibrium populations, we first calculate the downwards rate $k_{1\leftarrow2}$ and then impose the detailed-balance condition based on
\begin{equation}
\frac{k_{b\leftarrow a}}{k_{a\leftarrow b}}
=
\frac{
\mathcal P^{(\mathrm a)}_{b,\mathrm{eq}}
}{
\mathcal P^{(\mathrm a)}_{a,\mathrm{eq}}
}.
\end{equation}
Since the nuclear degrees of freedom are treated classically, we calculate the
adiabatic equilibrium population using the single-bead, classical-nuclear limit,
\begin{equation}
   \mathcal{P}^{(\rm a)}_{a,{\rm eq}}
   =
   \frac{
   \int \mathrm{d}\mathbf Q \,
   e^{-\beta U_{\rm RP}(\mathbf Q)}
   \bra{\phi_a(\mathbf Q)}
   e^{-\beta \hat U_{\rm ex}(\mathbf Q)}
   \ket{\phi_a(\mathbf Q)}
   }{
   \int \mathrm{d}\mathbf Q \,
   e^{-\beta U_{\rm RP}(\mathbf Q)}
   \operatorname{Tr}_{\rm ex}
   \left[
   e^{-\beta \hat U_{\rm ex}(\mathbf Q)}
   \right]
   } .
\end{equation}

\bibliography{main}